\documentclass[fleqn]{aa}
\usepackage{graphicx}
\usepackage[varg]{txfonts}           
\usepackage{natbib}           

\bibpunct{(}{)}{;}{a}{}{,}
\defcitealias{Schetal.05}{Paper II}
\defcitealias{Pe.04}{Paper I}
\defcitealias{Schetal.05a}{Paper III}

\newcommand{\kms}{km\,s$^{-1}$}
\newcommand{\Msun}{M$_{\sun}$}

\newcommand{\Lsun}{L$_{\sun}$}
\newcommand{\lsun}{L_{\sun}}
\newcommand{\Mdot}{M$_{\sun}$\,yr$^{-1}$}

\newcommand{\etal}{et al.}
\newcommand{\hb}{H$\beta$}
\newcommand{\ha}{H$\alpha$}
\newcommand{\heii}{He\,{\sc ii}}
\newcommand{\teff}{T_\mathrm{eff}}

\newcommand{\nii}{[N\,{\sc ii}]}
\newcommand{\oiii}{[O\,{\sc iii}]}

\begin{document}

\title{The evolution of planetary nebulae
          }

   \subtitle{IV. On the physics of the luminosity function\thanks{
             Based in parts on observations
             made with the NASA/ESA Hubble Space Telescope, obtained at the
	     Space Science Institute, which is operated by the Association of the
	     Universities for Research in Astronomy, Inc., under NASA contract
	     NAS 5-26555. The data are retrieved from the ESO/ST-ECF Science Archive
	     Facility.}
             }

   \author{
          D. Sch\"onberner
          \and
          R. Jacob
          \and
          M. Steffen
	  \and
	  C. Sandin
	  }

   \offprints{D. Sch\"onberner}

   \institute{
              Astrophysikalisches Institut Potsdam,
              An der Sternwarte 16, 14482 Potsdam, Germany \\
              \email{deschoenberner@aip.de, msteffen@aip.de}
	     }

   \date{Received .....; accepted .....}

\abstract
   {The luminosity function of planetary nebulae, in use for about two
   decades in extragalactic distance determinations, is still subject to
   controversial interpretations.}
    {The physical basis of the luminosity function is investigated by means
    of several evolutionary sequences of model planetary nebulae computed with a 1D
    radiation-hydrodynamics code.}
    {The nebular evolution is followed from the vicinity of the asymptotic-giant
     branch across the Hertzsprung-Russell diagram until the white-dwarf
     domain is reached, using various central-star models coupled to different
     initial envelope configurations.  Along each sequence the
     relevant line emissions of the nebulae are computed and analysed.}
    {Maximum line luminosities in \hb\ and [\ion{O}{iii}] 5007~\AA\
    are achieved at stellar effective temperatures of about 65\,000~K and 
    95\,000\dots100\,000~K, respectively, provided the nebula remains optically thick for
    ionising photons.   In the optically thin case, the maximum line
    emission occurs at or shortly after the thick/thin transition.
    Our models suggest that most planetary nebulae with hotter
    ($\ga 45\,000$~K) central stars are optically thin in the Lyman continuum,
    and that their [\ion{O}{iii}] 5007~\AA\ emission fails to explain the
    bright end of the observed planetary nebulae luminosity function.
    However, sequences with central stars of  $\ga\!0.6$~\Msun\ and
    rather dense initial envelopes remain virtually optically thick
    and are able to populate the bright end of the luminosity function.
    Individual luminosity functions depend strongly on the central-star
    mass and on the variation of the nebular optical depth with time.}
   {Hydrodynamical simulations of planetary nebulae are essential for any
    understanding of the basic physics behind their observed luminosity function.
    In particular, our models do not support the claim of 
    Marigo \etal\ (2004) according to which the maximum 5007~\AA\
    luminosity occurs during the recombination phase well
    beyond 100\,000~K when the stellar luminosity declines and the
    nebular models become, at least partially, optically thick.
    Consequently, there is no need to invoke relatively massive
    central stars of, say $>\! 0.7$~\Msun, to account
    for the bright end of the luminosity function.}


\titlerunning{The evolution of planetary nebulae IV.}
\maketitle

\section{Introduction}                                   \label{intro}
   Since the pioneering paper by \citet{J.89} in which the theoretical fundament
   of the planetary nebulae luminosity function (PNLF) has been
   laid out, the use of this tool for establishing cosmic distances has been
   proven to be extremely successful. Its main success stems from the fact that
   it works not only for spirals, but also for ellipticals, despite the
   fact that both systems consists of completely different stellar populations.
   A recent summary of the use of the PNLF can be found in \citet[][and
   references therein]{C.03}.

  \citet{J.89} defined the line magnitudes as
\begin{equation}
 m = -2.5\,\log F - 13.74\, ,
\end{equation}
  with $F$ (in erg\,cm$^{-2}$\,s$^{-1}$) being the line flux from the object.
  Based on 13 galaxies with
  well-determined Cepheid distances, an absolute cut-off brightness of
  $M^{\star}(5007)= -4.45\pm0.05$\, mag has been derived \citep{C.03}.  This brightness
  corresponds to 620~\Lsun\ emitted in the \oiii\ 5007\,\AA\ line!

   Despite its use for about two decades, the physical basis of the PNLF
   is still mysterious and subject to controversal interpretations.  A major
   uncertainty is related to the question whether the brightest planetary nebulae
   (PNe) are optically thick or thin for Lyman continuum photons because the
   efficiency of converting stellar UV radiation into optical line emission is
   heavily dependent on the optical depth.

   \citet{J.89} considered only optically thick nebular models for the
   construction of a theoretical PNLF.  Based on the post-asymptotic giant
   branch (post-AGB) tracks available, \citeauthor{J.89} computed the line
   luminosities from expanding filled spheres along tracks with
   hydrogen-burning and helium-burning central stars, and found, by comparing
   the brightest PNe in Local Group galaxies with these models, upper mass
   limits for central stars of $\approx\! 0.65$~\Msun.

   Also in the work of \citet{DJV.92} for Magellanic Cloud PNe only optically
   thick nebular models were considered.  These authors found that the maximum
   stellar luminosity belonging to the \oiii\ cutoff is $\simeq\! 10^4$~\Lsun,
   corresponding to $\simeq\! 0.7$~\Msun\ if the standard core-mass luminosity
   relation of hydrogen-burning central stars is used.

   \citet{Stangh.95} modelled only the H$\beta$\ luminosity function, assuming also
   optically thick nebulae, and investigated the dependence on star formation rate,
   post-AGB transition time and initial-final mass relation.   The observed
   H$\beta$\ cut-off at $\simeq\!-2.5$~mag was achieved by nebular models around nuclei
   with 0.65~\Msun\ for a fairly broad range of assumptions.

   Considering only models that are optically thick to ionising radiation
   oversimplifies the problem certainly.
   Thus \citet{MKCJ.93} and \citet{MS.97} used a different method. They
   modelled luminosity functions based on hydrogen-burning post-AGB
   evolutionary tracks and empirical properties of PNe \emph{without} resorting
   to nebular models.  Their main conclusion is that the bright end
   of the luminosity function is predominantly populated by optically thin PNe
   (not thin in all directions around the central star, but thin in at least
   some directions to allow for some leaking of Lyman continuum photons)
   with maximum central-star masses between 0.63 and 0.66~\Msun.

   A completely novel approach was introduced by \citet{Ma.01,Ma.04}.
   These authors constructed simple nebular models, taking into
   account their outer boundary conditions set by the AGB and the central-star
   winds, using analytical expressions for interacting winds in a similar way
   as described by \citet{VK.85}.   The pressure increase within the nebular shell
   by photoionisation is approximately considered.  This new
   ``synthetic" approach allows to compute the evolution of model PNe together
   with their observable quantities in a very fast and efficient way.

   Using their new tool, \citet{Ma.04} computed nebular sequences along the
   post-AGB tracks of \citet{VW.94} and constructed PNLFs for stellar populations
   with various metallicities and star formation histories.  It turned out that
   the bright end of the PNLF is populated by objects with central-star masses
   between 0.70 and 0.75~\Msun.   Consequently, the value of the bright cut-off of
   the PNLF must depend critically on the properties of the parent stellar
   population: large differences between the maximum PNe luminosities of
   elliptical and spiral galaxies are to be expected
   \citep[see][Fig.~25 therein]{Ma.04}, which, however, are not observed.

   This completely different interpretation of the luminosity function prompted
   us to employ our hydrodynamical models presented in
   \citet[][Paper I hereafter]{Pe.04} to
   investigate the physical basis of the PNLF with a more realistic
   approach.   We believe that hydrodynamical simulations are ideally suited
   to tackle this task since it has been demonstrated on several occasions
   that a reasonably good match to observed properties of PNe is achieved
   with such models, provided appropriate initial and boundary conditions
   are chosen \citep{Schetal.97, Schetal.99, Schetal.03a, Schetal.03b, SS.06}.
   However, we do not aim at constructing a theoretical luminosity function
   because the number of available combinations of nebular models and
   central-star masses is too low.   Instead, we will concentrate on a
   description of the processes responsible for the strength of the
   relevant line emissions.

   We begin in Sect.~\ref{mod.PN} with a short description of our hydrodynamical
   modelling including a brief description of the evolutionary properties of PNe
   as they follow from these models.
   We outline in Sect.~\ref{com.marigo} the basic differences to the Marigo \etal\
   approach and investigate in Sect.~\ref{line.emission} in detail how the
   luminosities of important lines depend on the model properties and how
   they evolve with time.   Section~\ref{LF}
   is devoted to individual luminosity functions as predicted by our
   simulations.  Section \ref{hbeta.pnlf} introduces the corresponding \hb\
   luminosity functions. The paper is closed by Sect.~\ref{dis} with an extensive
   discussion. A short presentation of the basic results of this investigation has
   already been given by \citep{Schetal.06}.

\section{Modelling the planetary nebulae evolution}          \label{mod.PN}
   A detailed description how we simulate the formation and evolution
   of planetary nebulae has already been given in
   previous papers and shall not be repeated here \citep{Pe.98,Pe.04}.
   Instead, we emphasise here only the basic ingredients of our calculations
   which are important for the general appreciation of our models and for
   understanding the differences to the ``synthetic" approach of \citet{Ma.01}.

   In this context it is important to remember that a planetary nebula is a very
   complex dynamical system, even if we approximate real objects by spherical
   configurations.   The whole object consists of a rapidly evolving post-AGB star
   and an expanding circumstellar envelope originally set up by a massive
   stellar wind when the object was still on or close to the AGB.
   The radiation field and the wind from the central star initiate a shock wave
   pattern at the inner edge of the slowly expanding AGB wind envelope.  Its
   structure and expansion properties depend mainly on the radial density distribution
   of the AGB material, on the electron temperature inside the ionised matter,
   and on the pressure support from the central-star wind.  The PN proper is
   confined between an inner contact surface which separates the nebula matter
   from the shock-heated wind matter, and an outer shock front which
   propagates through the ambient AGB material, thereby increasing the
   PN mass with time.

   \emph{Any approach with the aim to understand at least the basic physics
   of the formation and evolution of PNe {must} therefore rely on
   radiation-hydrodynamics simulations with the proper initial and boundary
   conditions, with all the relevant physical processes treated fully
   time-dependently.}

\subsection{The hydrodynamical models}                       \label{hydro.mod}
   In short, the basic philosophy of our hydrodynamical PN models is
   to couple a spherical circumstellar envelope, assumed to be the relic of
   a strong AGB wind, to a post-AGB model and to follow the evolution of
   the whole system across the Hertzsprung-Russell diagram towards the white-dwarf
   cooling path, employing an 1D radiation-hydrodynamics code
   \citep[see][]{Pe.98}.  We emphasize that our code is designed to compute
   ionisation, recombination, heating, and cooling fully time-dependently.
   For each volume element, the cooling function  is composed of the
   contributions of all the ions considered (next to H and He, also C, N, O,
   Ne, S, Cl, and Ar).  For each individual element up to 12 ionisation stages
   are taken into account.  More details on the radiation part of our code can be
   found in \citet{MaS.97}.

   We used in all of our computations a chemical composition typical for
   Galactic disk PNe (Table\,\ref{tab.element}).  Although the line emission of
   a PN and the luminosity and radiation field of its central star depend on the
   metal content, the effect is relatively small and will not influence the basic
   properties of the PNLF \citep[cf.][]{DJV.92}.   Any abundance variations,
   notably that of oxygen, are thus not considered in the present work.

\begin{table}[t]           
\caption{Elemental abundances, $\epsilon_\mathrm{i}$, used in the computations of
         our hydrodynamical models, in (logarithmic) number fractions relative to
	 hydrogen,
	 $\log {\epsilon}_\mathrm{i} = \log n_\mathrm{i}/\log n_\mathrm{H} + 12$.}
\label{tab.element}
\centering
\begin{tabular}{ccccccccc}
\hline\hline\noalign{\smallskip}
  H    &    He  &   C    &   N   &    O   &   Ne   &    S   &  Cl    &  Ar  \\
\noalign{\smallskip}\hline\noalign{\smallskip}
 12.00 &  11.04 &  8.89  &  8.39 &  8.65  &  8.01  &  7.04  &  5.32  &  6.46 \\
\noalign{\smallskip}\hline
\end{tabular}
\end{table}

   The hydrodynamical model sequences selected for the present work
   from Paper I are listed in Table\,\ref{tab.mod}.   The table provides the
   sequence numbers from Paper I (col.~1), the central-star masses
   (col.~2), the central-star luminosities at 30\,000~K (col.~3),
   the AGB mass-loss rates (col.~4) and the AGB wind velocities
   (col.~5) of the initial models, and the envelope types (col.~6).
   The initial envelope models have either an ad hoc radial power-law
   density distribution, $\rho \propto r^{-2}$
   (\textsc{Type A}), or are the result of detailed hydrodynamics simulations
   along the upper AGB (\textsc{Type~C}), as described in \citet{Schetal.97}
   or in Paper I.  The structure of the \textsc{Type~C} envelope reflects the
   mass-loss history of the preceding 50\,000 years of AGB evolution while
   \textsc{Type~A} corresponds to the simple case of constant mass loss and
   outflow velocity.

   An additional stellar track was introduced in
   \citet[][Paper II hereafter]{Schetal.05} by replacing the 0.605~\Msun\ track
   of sequence No.~6 by a 0.595~\Msun\ track interpolated from the 0.605~\Msun\
   and 0.565~\Msun\ tracks of \citet{B.95b} and \citet{Sch.83}.
   The PNe models of this new sequence (No.~6a in Table~\ref{tab.mod}) match
   the observations even better than the models of sequence No.~6 with a
   0.605~\Msun\ central star \citep[][Paper III hereafter]{Schetal.05a}.

\begin{table}[!t]                
\centering
\caption{Hydrodynamical sequences of model planetary nebulae
         used in this work. The sequence numbers (col. 1) refer to Table\,1 in Paper~I.
         Central star parameters at \hbox{$\teff=30\,000$} K are given in cols.~(2)
	 and (3), and cols.~(4) and (5) indicate the AGB mass-loss parameters.
         The peak mass-loss rate of the hydrodynamical simulation is about
	 $1\times10^{-4}$~\Mdot.
}
\label{tab.mod}
\begin{tabular}{rcrccc}
\hline\hline\noalign{\smallskip}
No. & $M $    & $L~~~$        & $\dot{M}_{\rm agb}$       & $V_{\rm agb}$ & {\sc Type}   \\
\noalign{\smallskip}
    & [\Msun] & [\Lsun]    & [\Mdot]          &        [\kms]          &       \\
\noalign{\smallskip}
(1) &   (2)   &    (3)~~     &      (4)         &       (5)              &  (6)  \\
\noalign{\smallskip}\hline\noalign{\smallskip}
   22    &  0.565& 3\,883  & $ 3\times10^{-5} $         &  10           & {\sc A}\\
\noalign{\medskip}
    4    &  0.605& 6\,280  & $ 1\times10^{-4} $         &  10           & {\sc A}\\
    6 & 0.605& 6\,280 & Hydro.\ Simulation &\hspace*{-2.5mm}$\simeq\! 12$ & {\sc C}\\
  6\,a& 0.595& 5\,593 & Hydro.\ Simulation &\hspace*{-2.5mm}$\simeq\! 12$ & {\sc C}\\
\noalign{\medskip}
    8    &  0.625& 7\,900  & $ 1\times10^{-4} $         &  15           & {\sc A}\\
\noalign{\medskip}
   10    &  0.696&11\,615  & $ 1\times10^{-4} $         &  15           & {\sc A}\\
\noalign{\smallskip}\hline
\end{tabular}
\end{table}

  For the purpose of this work we recalculated the sequences listed in
  Table~\ref{tab.mod} by means of an updated version of our radiation-hydrodynamics
  code.  In particular, the radiation transport is now treated in the
  `outward only' approximation, and heat conduction as described in
  \citet{SchSt.06} is also included.  These improvements of the physics did
  not lead to any significant changes of the dynamical structures of the PN models.
  Thus all the conclusion obtained in
  earlier publications and which are based on the older simulations remain valid.

   All the central-star models used in our PNe simulations are burning hydrogen,
   and they are assumed to radiate as black bodies.
   This choice is justified since \citet{GKM.91} showed by means of
   so-called Unified NLTE model atmospheres that a
   black-body energy distribution with the effective temperature of the
   photosphere provides a good empirical description of the stellar
   UV flux.   This holds at least for effective temperatures between approximately
   40\,000 and 100\,000~K.

\begin{figure*}[!ht]          
\sidecaption
\includegraphics[width= 13.5cm]{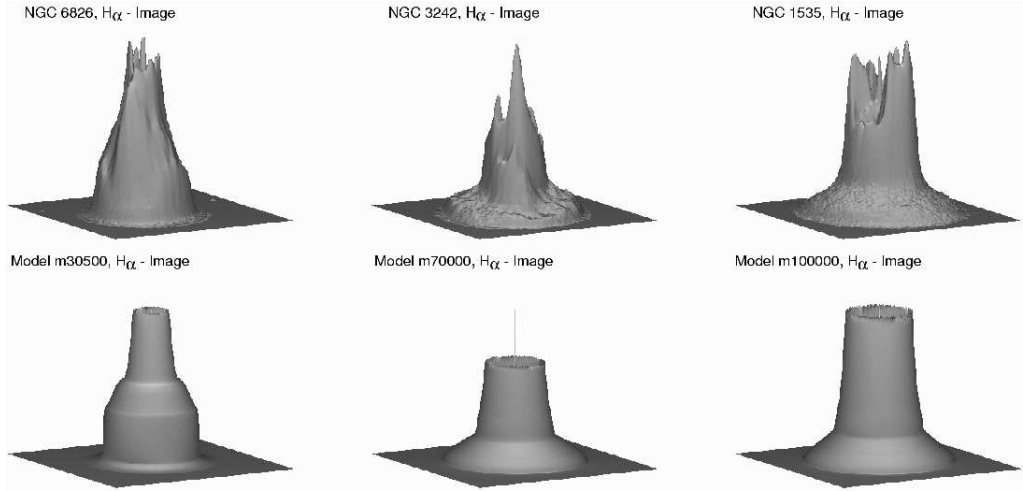}
\caption{\emph{Top}:  normalized 3D representations of the \ha\ images
                      of \object{NGC~6826} (\emph{left}), \object{NGC~3242}
		      (\emph{middle}), and \object{NGC~1535} (\emph{right}).
		      The images are from an unpublished catalogue compiled by
		      B\"assgen (priv. communication).
        \emph{Bottom}: corresponding \ha\ surface brightness profiles of optically
	              thin models selected from the hydrodynamic sequence No.~6 listed
		      in Table~\ref{tab.mod}.   Post-AGB ages increase from left
		      to right.  The normalizations of the models are adjusted
		      as to match those of the observed objects.
        }
\label{bottle.pn.model}
\end{figure*}

  The success of these detailed radiation-hydrodynamics simulations in describing
  planetary nebulae is illustrated in Fig.\,\ref{bottle.pn.model} where
  we compare the \ha\ brightness distributions of 3 PNe with
  well-developed double-shell structures with the model predictions.
  The models are selected from sequence No.\,6 of Table~\ref{tab.mod} which
  started with an initial envelope computed by means of two-component
  radiation-hydrodynamics simulations along the upper AGB
  \citep[{\sc Type C}, see][for more details]{Stefetal.98}.

   Although the models are spherically symmetric,
  they represent the observed general structures as indicated by the H$\alpha$\
  brightness distribution astoundingly well:
   the brightness ratio between both shells, i.e.\ `rim'
  and `shell', is well matched, and also the linear brightness slope of faint
  `shells', typical for many nebulae, is reproduced.  Such a linear radial
  brightness profile of the `shell' develops if the radial density gradient
  of the circumstellar envelope steepens with distance from the star,
  indicative of increasing mass loss towards the end of the AGB evolution
  \citep{Stefetal.98, Schetal.05}.   More comparisons between observed structures
  of planetary nebulae and the predictions of our radiation-hydrodynamics
  simulations can be found in \citet{SS.06}.

\subsection{General behaviour of our nebular models}         \label{mod.behave}

  For a better appreciation of the differences between our hydrodynamical models
  and those developed by \citet{Ma.01} it appears to be useful to give a brief
  description of the basic principles of the formation and evolution of
  planetary nebulae as they follow from recent realistic hydrodynamics simulations.
  In Paper~I we introduced several phases of the PN evolution according to the
  main physical processes acting on the whole system.  We briefly repeat them here,
  using the models of sequence No.~6 for illustration.

\begin{figure}[!t]
\includegraphics[width=1.0\columnwidth]{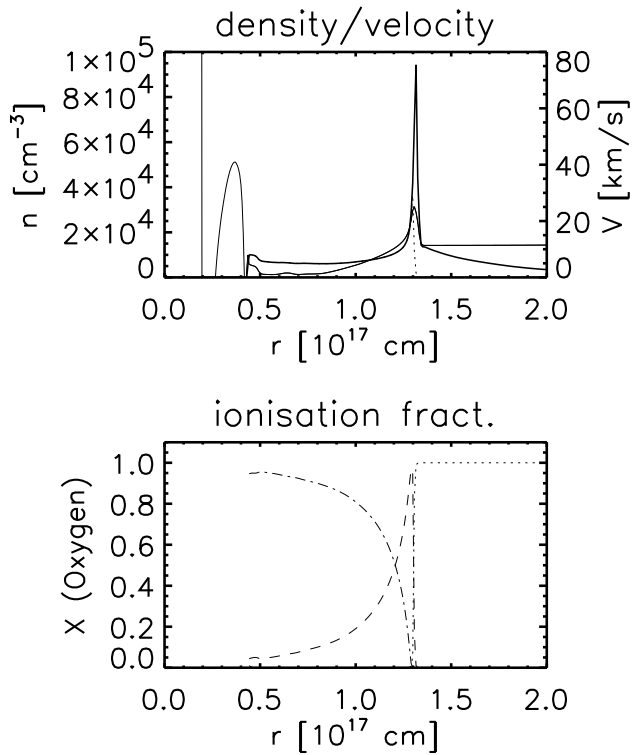}
\caption
   {\emph{Top}:  radial profiles of
   heavy-particle density (thick), electron density (dotted), and velocity (thin)
   of an optically thick model selected from sequence No.~6.
   The stellar parameters of the 0.605\,\Msun\ central star are {$\teff = 40\,245$}\,K
   and {$L = 6240$}~\Lsun, with a post-AGB age {$t= 2478$}\,yr.
   \emph{Bottom}: ionisation fractions of neutral (dotted), singly ionised (dashed),
   and doubly ionised oxygen (dash-dotted).  The ionisation fractions within the
   shocked central-star wind domain ($r\le 0.5\cdot10^{17}$~cm), though computed,
   are not shown for clarity.  
   }
\label{ion1}
\end{figure}

  Neglecting the proto-planetary-nebula phase which is of no interest here,
  the first important phase, the \textbf{ionisation phase}, begins when the
  intense flux of ionising photons from the central star starts to ionise and to
  heat the inner parts of the AGB wind envelope.    The large thermal pressure of
  the ionised gas determines shape and expansion of the newly created PN.
  The ionisation front is of type D and trapped by a strong, nearly isothermal
  shock (see Fig.~\ref{ion1}, top panel).

  The bottom panel of the figure depicts the ionisation structure of oxygen, which
  is of particular interest for the present study.  The degree of ionisation is still
  rather moderate: neutral in the undisturbed AGB wind, singly ionised preferently in
  the outer and doubly ionised in the inner parts of the ionised shell.

\begin{figure}[!t]
\includegraphics[width=1.0\columnwidth]{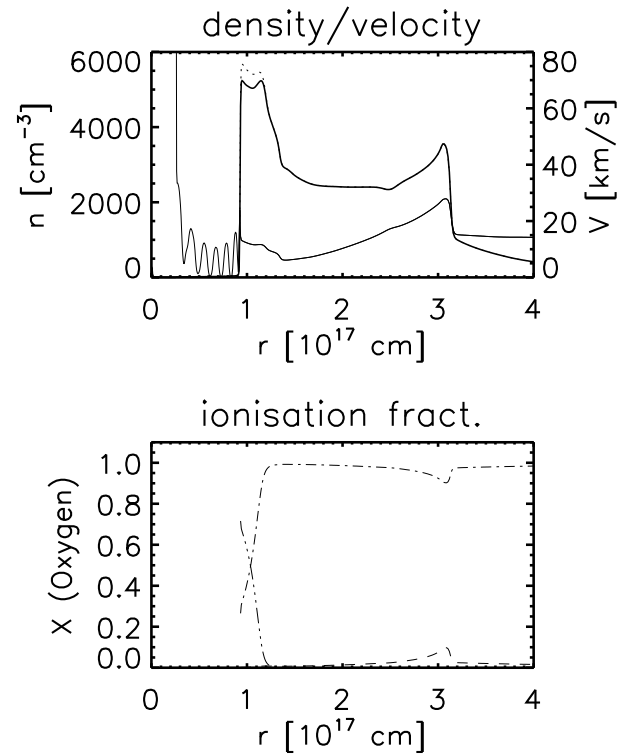}
\caption
   {Same as in Fig.~\ref{ion1}, but for an optically thin model with well developed
    rim and shell structures.  The stellar parameters are $\teff = 79\,708$\,K and
    $L = 5797$~\Lsun, with a post-AGB age $t= 4241$\,yr.  The dash-triply dotted
    line indicates triply ionised oxygen.
   }
\label{ion2}
\end{figure}

  The ionisation phase finishes if
  the ionisation front passes the shock and propagates quickly through
  the still undisturbed AGB matter (R-type ionisation front).   The PN is then
  optically thin for ionising radiation and said to be density bounded.
  Figure~\ref{ion2} depicts a moment well after
  the thick/thin transition.  The outer boundary of the PN is now the
  shock front enclosing the `shell', and not the ionisation front which has
  already left the computational domain.
  The main ionisation stage of oxygen is O$^{+2}$ throughout the shell and the
  halo, except in the inner part of the rim where we have already O$^{+3}$.
  Helium is also doubly ionised in this inner region, and the extra heat
  deposited by the ionisation of He$^+$ drives a weak shock
  (at $r=1.3\cdot10^{17}$~cm in Fig.~\ref{ion2}, top panel).

  The model PN is now in the \textbf{compression phase} in which the high
  pressure of the shock-heated central-star wind accelerates and compresses the inner
  parts of the shell into the so-called `rim'.
  The rim becomes the dominant structure of a PN, and the objects
  displayed in Fig.\,\ref{bottle.pn.model} are in this stage.   The compression
  phase is terminated when the wind power declines because the central star
  is approaching its maximum effective temperature and fades towards
  a white-dwarf configuration.   

\begin{figure}[!t]
\includegraphics[width=1.0\columnwidth]{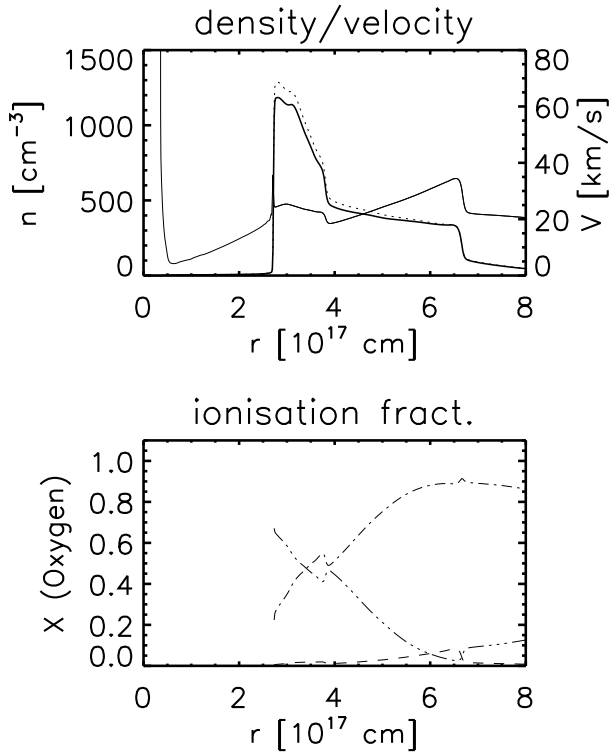}
\caption
   {Same as in Fig.~\ref{ion1}, but for an optically thin model at the end of the
   compression phase close to maximum stellar temperature.
   The stellar parameters are $\teff = 156\,662$\,K and $L = 2075$~\Lsun,
   with a post-AGB age $t= 7124$\,yr.
   }
\label{ion3}
\end{figure}

  Figure~\ref{ion3} shows a model very close to maximum stellar temperature at the
  end of the compression phase.  At the large effective temperature well above
  100\,000~K a significant fraction of oxygen is now triply ionised.  There
  exists, however, still some singly ionised oxygen close to the outer edge of the
  shell.

  During the whole compression phase the leading edge of the shell continues
  to propagate supersonically into the AGB wind with a (relative) speed ruled only
  by the thermal properties of the gas and the radial density profile of the AGB
  wind \citep[][]{FTTB.90, Ch.97, Shuetal.02, Schetal.05}.
  \emph{The expansion of the shell is not at all influenced by the wind
  interaction responsible for the formation of the rim.}  Usually, the shell
  expands faster than the rim \citepalias[see][Fig.\,12 therein]{Schetal.05}.

  The final fading of the central star causes
  \textbf{recombination} in the outer parts of the PN provided the nebular
  densities are sufficiently large, and/or the luminosity declines very rapidly.
  Recombination turns eventually into \textbf{re-ionisation} after the
  fading of the central star has slowed down and the nebular density becomes
  sufficiently low due to the continued expansion.

\begin{figure}[t]
\includegraphics[width=1.0\columnwidth]{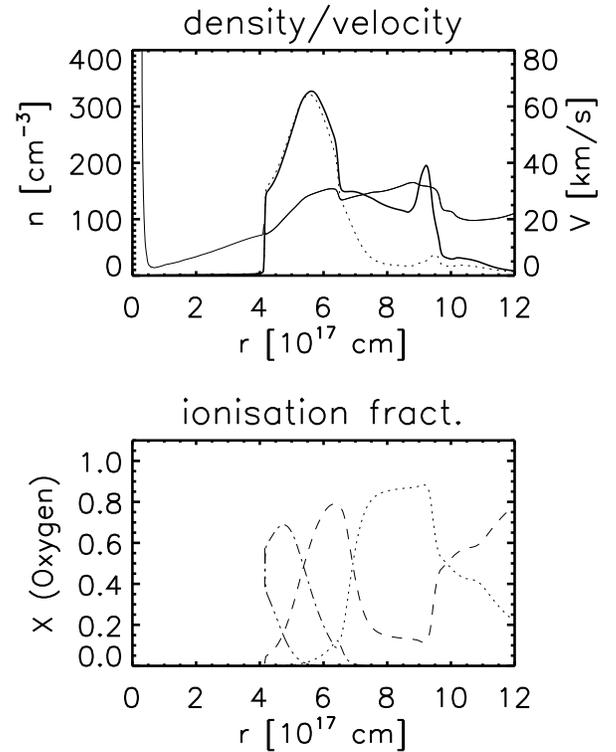}
\caption
   {Same as in Fig.~\ref{ion1}, but for a model after the end of recombination and at
   the beginning of re-ionisation.  The stellar parameters are {$\teff = 122\,269$}\,K
    and {$L = 249$}~\Lsun, with a post-AGB age $t= 9557$\,yr.
   }
\label{ion4}
\end{figure}

  The model shown in Fig.\,\ref{ion4} illustrates the situation after the end of
  the recombination and at the beginning of the re-ionisation stage.  The shell is
  mainly neutral, although not completely, and forms for a while a {\bf recombination halo}
  \citep{Coretal.00}.  The rim which remained fully ionised to a large extent
  continues to expand and will eventually swallow the shell material because the shell's
  shock is slowed down during recombination.
  The ionisation is highly stratified (bottom) and reflects the physical
  situation given by a central star of low luminosity but still very high effective
  temperature:  mainly O$^0$ in the shell, but the degree of ionisation increases
  inwards over O$^+$ until O$^{+3}$ at the inner rim.

  These main evolutionary stages as described here may not always occur,
  or they may even occur at the same time.  For instance, a PN around
  a low-mass, slowly evolving central star will not recombine at all.  On the
  other hand, a PN around a massive, very quickly evolving central star may not
  become optically thin.
  In this case the ionisation front remains always of type D, i.e.\ trapped
  by the outer shock, and we have ionisation and compression at the same time.
  However, the double shell configuration is very robust and develops also in
  these cases.

  This discussion shows that the formation of typical PN structures is quite
  complex even in the spherical approximation.   Two processes are relevant, viz.\
  heating by ionisation and compression by wind interaction.
  Their relative importance will depend on the
  metallicity, i.e. on the content of coolants.
  For instance, at lower metallicities than we have used here we expect (i)
  less dense rims simply because the central-star wind power is most likely
  lower, and (ii) higher expansion rates of the shell because the temperature,
  and sound speed, is larger.

  A completely different situation is encountered in systems with a Wolf-Rayet
  central star. The virtually hydrogen-free wind from such a star is up to two
  orders-of-magnitude more intense compared with objects of normal surface
  abundances \cite[cf.][]{Leuetal.96}.
  Consequently, wind interaction will always be dominant for shaping a PN
  around a Wolf-Rayet central star, independently of the general metallicity.
  Since origin and evolution of hydrogen-poor central stars is not known, nebular
  models around them can not be computed to date.   The sequences listed in
  Table~\ref{tab.mod} und used here have exclusively central-star models with normal
  surface composition burning hydrogen to provide their luminosity.

\section{Comparison with the Marigo \etal\ models}         \label{com.marigo}
  As already mentioned in the Introduction, the approach of \citet{Ma.01} to compute
  the evolution PNe analytically goes back to methods developed by \citet{VK.85}.
  Marigo \etal\ introduced several improvements, the most important one being the
  approximate consideration of radiative processes, viz.\ heating by ionisation and
  cooling by line emissions.   However, an analytical treatment of the dynamics of
  photoionisation oversimplifies the problem and has severe consequences.

  For instance, the basic nebular structures of the Marigo \etal\ models are
  {shells of constant density which are partially or fully ionised}.
  They are in full contrast to real objects
  which have complicated density and velocity profiles and which can only be approximated
  by hydrodynamical models (cf.\ Figs.~\ref{ion1} till \ref{ion4}).  In a typical
  double-shell PN most of the nebular mass ($\ga\! 80\,\%$) is contained in a shell of
  rather low density and expands \emph{independently} of the properties of the
  central-star wind, as explained in the previous section.  Only the small fraction of
  nebular matter contained in the rim is ruled by wind interaction.

  The density structure of a PN, however, is important for the ionisation balance
  since local photoionisation rates are proportional to density, recombination rates,
  however, proportional to density squared.  On one hand, the ionisation front may
  be kept trapped for quite a while during the early evolution when the nebular
  densities are still large (D-type front), resulting in a delay of the thick/thin
  transition.   On the other hand an extended, low density shell is less
  prone to recombination than a Str\"omgren sphere of the same mass but with larger,
  constant density.   The question of the nebula's optical thickness
  to ionising radiation, however, is crucial for the emission-line luminosities and
  for the interpretation of the luminosity function \citep[see][]{MKCJ.93,MS.97}.

  The influence of the different treatment of nebular shells becomes evident by
  a direct comparison between the properties of the hydrodynamical models with
  those of \citet{Ma.04}.   For this purpose,  we selected our sequence No. 6a
  together with the sequence No.\ 4 of \citet[][Table~2 and
  Fig.~9 therein]{Ma.01}\footnote
{Sequence No. 4 is the only one for which detailed information is available from
\citet[][Fig. 9 therein]{Ma.01}.}.
  Both simulations are based on central stars with very
  similar masses (0.595 vs.\ 0.5989~\Msun),  and also their
  final AGB mass-loss rates are about the same, $\approx\!10^{-4}$~\Mdot.
  Both sequences may be considered as typical representatives
  for the planetary nebula evolution.

  The Marigo model becomes optically thin
  in the Lyman continuum very early, at about 25\,000~K. It remains optically thin
  until the central star fades below 1000~\Lsun\ along the cooling part of the track
  where the shell recombines \citep[Fig.\,10 in][]{Ma.01}.
  In our model, the ionisation front remains trapped behind a strong shock until the
  central star becomes as hot as  $\teff \simeq 47\,000$~K.

  A direct observational test about the quality of PN simulations is possible
  by considering the (relative) \emph{geometrical} thickness of the whole nebular
  structure.  We define the total relative thickness of a PN  as
  $\delta R = (R_\mathrm{out} - R_\mathrm{cd})/R_\mathrm{out},$
  with $R_\mathrm{out}$ and $R_\mathrm{cd}$ being the radial positions of the
  outer nebular boundary and of the contact surface, resp.
 The nebular boundary is either defined by the ionisation front (optically
  thick case) or by the shell's shock for density bounded, optically thin models.
  During the recombination phase, the position of the recombination/ionisation
  front can not be determined precisely, nor is the recombination complete
  (see Fig. \ref{ion4}).
  We thus defined $R_\mathrm{out}$ in general by the radial position where 99\,\%
  of the total \hb\ line emission from the nebular model is included.

 The difference between $R_\mathrm{out}$ as defined above and the position of
  the shock, $R_\mathrm{shock}$, is negligible as long as the ionisation is complete.
  During recombination the outer parts of the shell may become faint to such an extent
  that $R_\mathrm{out}$ gets smaller than $R_\mathrm{shock}$.  For example, the model
  shown in Fig.~\ref{ion4} is at the end of its recombination phase, and we have
  $R_\mathrm{out} = 8.8\cdot10^{17}$ cm and $R_\mathrm{shock}=9.6\cdot10^{17}$ cm.

  Instead of plotting $\delta R$ vs. time or $R$, we used the stellar
  temperature as a distance-independent proxy for the post-AGB age
  because formation and evolution of the nebular structure is ruled by
  the radiation field and the wind from the central star, both of which depend
  directly or indirectly on the effective temperature.  The choice of the stellar
  temperature has the advantage that its range depends only modestly on
  stellar mass, much in contrast to the evolutionary time scale.

\begin{figure}[t]          
\includegraphics[bb= 0.2cm 0.5cm 19.7cm 14.6cm, width=\linewidth]{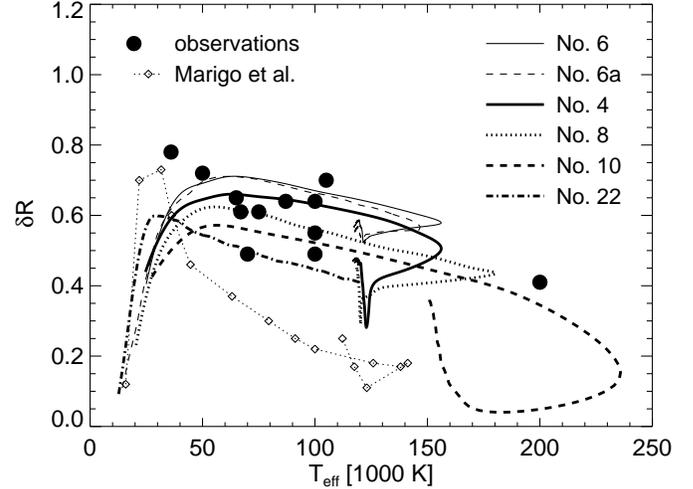}
\caption{Relative geometrical thickness,
        $\delta R = (R_\mathrm{out} - R_\mathrm{cd})/R_\mathrm{out}$, of PNe
	predicted by the different hydrodynamical sequences from Table
	\ref{tab.mod}, and by a sequence of \citet[][Figs.\ 9 \& 11 therein]{Ma.01}
	 vs.\ the  effective temperatures of the respective central stars, where the
	 temperatures are used as proxies for the post-AGB ages.
	 {The total times shown are 3\,000 yr for sequence No. 10, 10\,000 yr for No. 8,
	 about 15\,000 yr for the Marigo et al. sequence, and about 20\,000 yr for
	 the sequences Nos. 4, 6, and 6a.}
	 For comparison, the measured $\delta R$'s  of the PNe from Table~\ref{delta}
	 are plotted, too (filled circles).
	}
\label{rel.rad}
\end{figure}

  The variation of $\delta R$ with stellar effective temperature for {the
  hydrodynamical model sequences used here (see Table~\ref{tab.mod})} and for the
  Marigo sequence selected above is displayed in Fig.~\ref{rel.rad}.
  At low effective temperatures, the differences between the Marigo and our
  models are quite small, i.e. $\delta R$ increases rapidly to about 0.6.
  Then our models remain rather extended
  for the whole transition across the Hertzsprung-Russell diagram, while
  $\delta R$\ of the Marigo models decreases steadily from its maximum value
  of $\approx\!0.7$ to about only 0.2.  At this low value the model becomes
  optically thick by recombination \citep[cf.\ Fig.~9 in][]{Ma.01}.

 {Once the star has passed its maximum effective temperature, the outer
  regions of the hydrodynamical models may recombine more or less strong,
  leading to a rapid reduction of $\delta R$ to values as small as $\simeq\! 0.05$
  for the above adopted 99\% criterium for the determination of $R_\mathrm{out}$.
  Re-ionisation finally brings $\delta R$ up again.}

  The different model predictions concerning the geometrical structure of PNe
  can easily be checked against observations.  However,
  instead of using data from the literature, we preferred to remeasure the
  thicknesses of a number of well-known PNe whose (monochromatic)
  images were at our disposal, viz.\ the objects discussed in
  \citetalias[][]{Schetal.05} and \citetalias[][]{Schetal.05a}.
  While the outer edge of a PN, i.e.\ the location of the shell's shock,
  is easy to measure, the position of the inner edge, given by the contact surface,
  is not so obvious.   However, the surface brightness of nebular models
  reveal that the peak brightness of the rim in \oiii\
  (or in \ha) indicates the radial position of the contact surface.
  Guided by the models, we determined the relative thicknesses, $\delta R$,
  along the semi-minor axes by means of surface brightness plots derived from the
  monochromatic images and present the results in Table~\ref{delta}.

\begin{table}[t]                 
\caption{Relative geometrical thicknesses,
       $\delta R = (R_\mathrm{out} - R_\mathrm{cd})/R_\mathrm{out}$,
        measured along the semi-minor axis for a sample of well-observed PNe.
	The images used were either retrieved from the HST data archive, or they were
	taken with ground-based telecopes \citetext{Corradi, priv.\ comm.}.
        }
\label{delta}
\begin{center}
\begin{tabular}{lrcl}
\hline\hline\noalign{\smallskip}
  Object    &  $\teff$~[K]  & $\delta R$ &   Comments         \\
\noalign{\smallskip}
\hline\noalign{\smallskip}
  IC 418    &  36\,000         &   0.78     &    HST          \\
  IC 2448   &  65\,000         &   0.65     &    HST          \\
  NGC 1535  &  70\,000         &   0.64     &    Ground based \\
  NGC 2022  & 100\,000         &   0.55     &    Ground based \\
  NGC 2610  & 100\,000         &   0.49     &    Ground based \\
  NGC 3242  &  75\,000         &   0.61     &    HST          \\
  NGC 6578  &  67\,000         &   0.61     &    HST          \\
  NGC 6826  &  50\,000         &   0.72     &    HST          \\
  NGC 6884  &  87\,000         &   0.64     &    HST          \\
  NGC 7027  &$\simeq\!200\,000$&   0.41     &    HST\,$^\star$    \\
  NGC 7662  & 100\,000         &   0.64     &    HST          \\
  My 60     & 105\,000         &   0.70     &    Ground based \\[0.5mm]
\hline
\\[-6mm]
\end{tabular}
\end{center}
$^\star$ Semi-major axis (cf. Paper III).
\end{table}


  Our measured values listed in Table~\ref{delta} cluster around
  \mbox{$\delta \approx 0.6$} and are thus substantially larger than claimed
  by \citet{ZK.98} who found {typical values between 0.3 and 0.6}.
  However,  these authors considered only the bright nebular structures,
  viz.\ the rim in the cases of double shell PNe, ignoring the fact
  that the major part of the nebular mass is contained in the shell.
  {Using the data of \citet{ZK.98} we found an average value of only
  $\delta R\simeq 0.40$ for the 7 objects in common}.
  Inclusion of the shells would increase their (relative) thicknesses to our values.

  Figure~\ref{rel.rad} indicates that the relative thickness of PNe, once
  established by photoionisation during the early evolution,
  decreases only slightly during the
  further evolution across the Hertzsprung-Russell diagram, if at all.
  Our hydrodynamical models are in excellent agreement with the observations,
  contrary to the presented sequence of \citet{Ma.01} whose models become
  geometrically too thin during the course of their evolution.
 The reason for this difference is the behaviour of the propagation velocities
  of the outer shock and the contact discontinuity in the hydrodynamic approach:
  both are accelerated such that the relative distance between them does not change
  much.  For a detailed discussion of the expansion properties
  of hydrodynamical models see \citetalias{Schetal.05} and \citetalias{Schetal.05a}.
  We emphasize that our sequences shown in Fig.~\ref{rel.rad} span
  a range of final masses between 0.57 and 0.7 \Msun\ with the corresponding
  spread of initial masses.  Also, \emph{all} the observed objects  shown in the figure
  are on the high-luminosity part of their evolution to the white-dwarf domain.

  As can be seen from Fig.\,17 in \citet{Ma.01}, all their models
  behave like the one shown in Fig.~\ref{rel.rad}, i.e.\ $\delta R$ attains very
  low values around or even below 0.2 {in the hot
  region of the Hertzsprung-Russell diagram where the evolution is slowed down shortly
  before the maximum effective temperature of the central star is reached}\footnote
{\citet{Ma.01} use the nebular size as proxy of the post-AGB age, introducing thereby
  distance uncertainties into the comparisons with observations.}.
  {Yet the \cite{ZK.98} sample contains only very few objects with
  $\delta R \la 0.3$.  Their whole sample can only be reproduced by a superposition of
  sequences with different evolutionary time scales, a fact that does not seem to be
  very likely}.
  {We iterate that the real thickness of most objects from the \citeauthor{ZK.98}
  sample shown in Fig. 17 of
  \citet{Ma.01} is certainly larger, making it even more difficult to accept such
  a comparison as support for the Marigo et al. models}.

  \cite{Ma.01} conducted several other tests to verify the usefulness of their
   models.  Unfortunately, these tests suffer from inaccurate and inhomogeneous data
   and/or only poorly known distances.  This concerns diagrams like ionised mass vs.
   nebular size and electron density vs. nebular size.  Our sequences passed these
   tests equally well.

  A more specific comment on the use of expansion velocity--radius diagrams
   appears to be in order.  
   It has been shown by means of hydrodynamical simulations that the term `expansion
   velocity' has to be treated with utmost care \citepalias[cf.][]{Schetal.05, Schetal.05a}.
   The expansion measured by Doppler-split lines refers to a typical flow speed
   \emph{within} the nebular structure, while the
   real nebular expansion is given by the propagation of the shell's shock front.  A diagram
   like the one shown in Fig. 21 of \citet{Ma.01} where (observed) flow velocities are
   compared with boundary velocities or averages of them is certainly problematic.

   An additional uncertainty is being introduced by comparing model velocities
   with the \citet{We.89} compilation of expansion velocities.
   This data set is extremely inhomogeneous,
   and the entries refer mainly to the bright parts of the PNe which dominate the
   line emission and which are by no means representative for the real nebular
   expansion \citep[cf.][]{Getal.96}. In \citetalias{Schetal.05} we demonstrated that,
   once the initial conditions for the modelling are properly chosen and the
   expansion of objects correctly measured, hydrodynamical models give a reasonable
   description of the expansion properties of those planetary nebulae which do not
   differ too much from a spherical/elliptical shape.

  A distance independent observational test of relevance is the one
  on the typical structures of PNe conducted above.
  Real PNe are obviously not very compressed, and their outer parts
  do not recombine totally during the later phases of their evolution.
  The reason for the larger extent of real objects (and of our models) is
  the generation of a shock wave by photoionisation whose propagation
  is ruled by the electron temperature and the density gradient of the
  circumstellar matter and cannot be derived by applying energy and momentum
  balances based on a wind interaction scenario.
  It is the different physical description that obviously leads to the
  structural differences between our hydrodynamical and the \citet{Ma.01} models.

   {Based on all these facts we believe that our hydrodynamical models are better
    representatives of real planetary nebulae than the \citeauthor{Ma.01} models.
    Further support for this statement is presented in Sect.~\ref{excite}.

\begin{figure}[t]          
\includegraphics[width=1.\columnwidth]{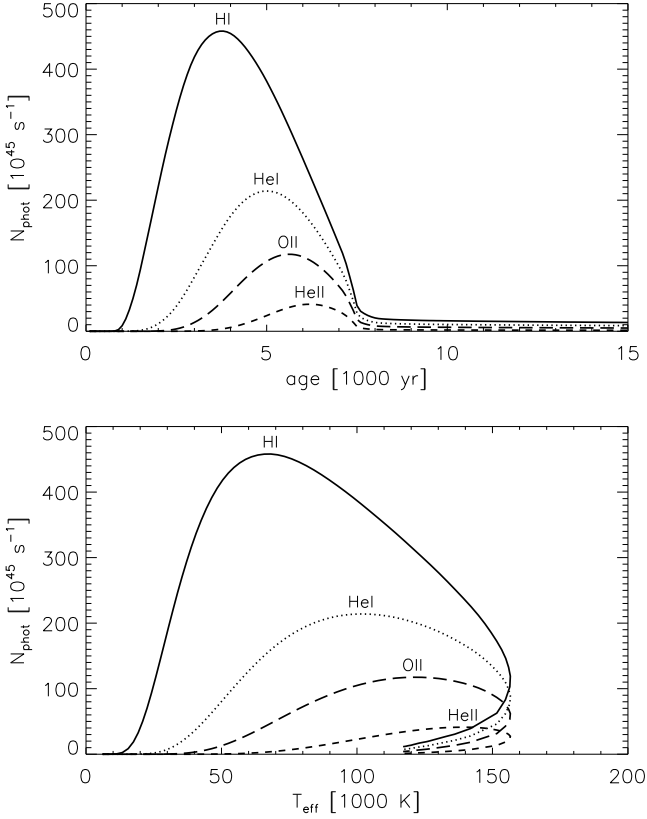}
\caption{Number of ionising photons emitted per second by a 0.605~\Msun\ post-AGB model
         as a function of age (\emph{top}) and effective temperature (\emph{bottom}),
	 under the assumption that the stellar photosphere radiates as a black body.
	 Plotted are the relations for hydrogen ($\lambda < 912$~\AA), neutral helium
	 ($\lambda < 504$~\AA) and singly ionised helium ($\lambda < 228$~\AA),
	 and for singly ionised oxygen ($\lambda < 353$~\AA).
}
\label{photon.605}
\end{figure}

\section{Evolution of emission lines}
\label{line.emission}
  With our hydrodynamical models we have the means to investigate in a more realistic
  way how the line emission depends on the central star's luminosity and
  effective temperature, and also on the nebula's optical depth, as the whole
  system evolves across the Hertzsprung-Russell diagram.
  Previous attempts to follow the line emission along an evolutionary track are
  based on purely kinematic models with simple geometries, constant densities,
  and ionisation equilibrium \citep[e.g.][]{Vi.83, Sta.89}.

  The line emission from a PN is, as is well known, determined by the
  ionising photon flux from the central star and the optical depth of the nebular
  shell.  Figure~\ref{photon.605} illustrates the run of the number of ionising
  photons emitted per second with age (top) and effective temperature (bottom) for
  the two 0.605\,\Msun\ sequences (Nos.~4 and 6).  Note that these total photon numbers
  depend not only on the effective temperature but also on the stellar radius, which
  explains the large variations with time or temperature.

\begin{figure}[t]          
\includegraphics[bb= 0.5cm 0.5cm 19.5cm 14.5cm, width=0.99\linewidth]{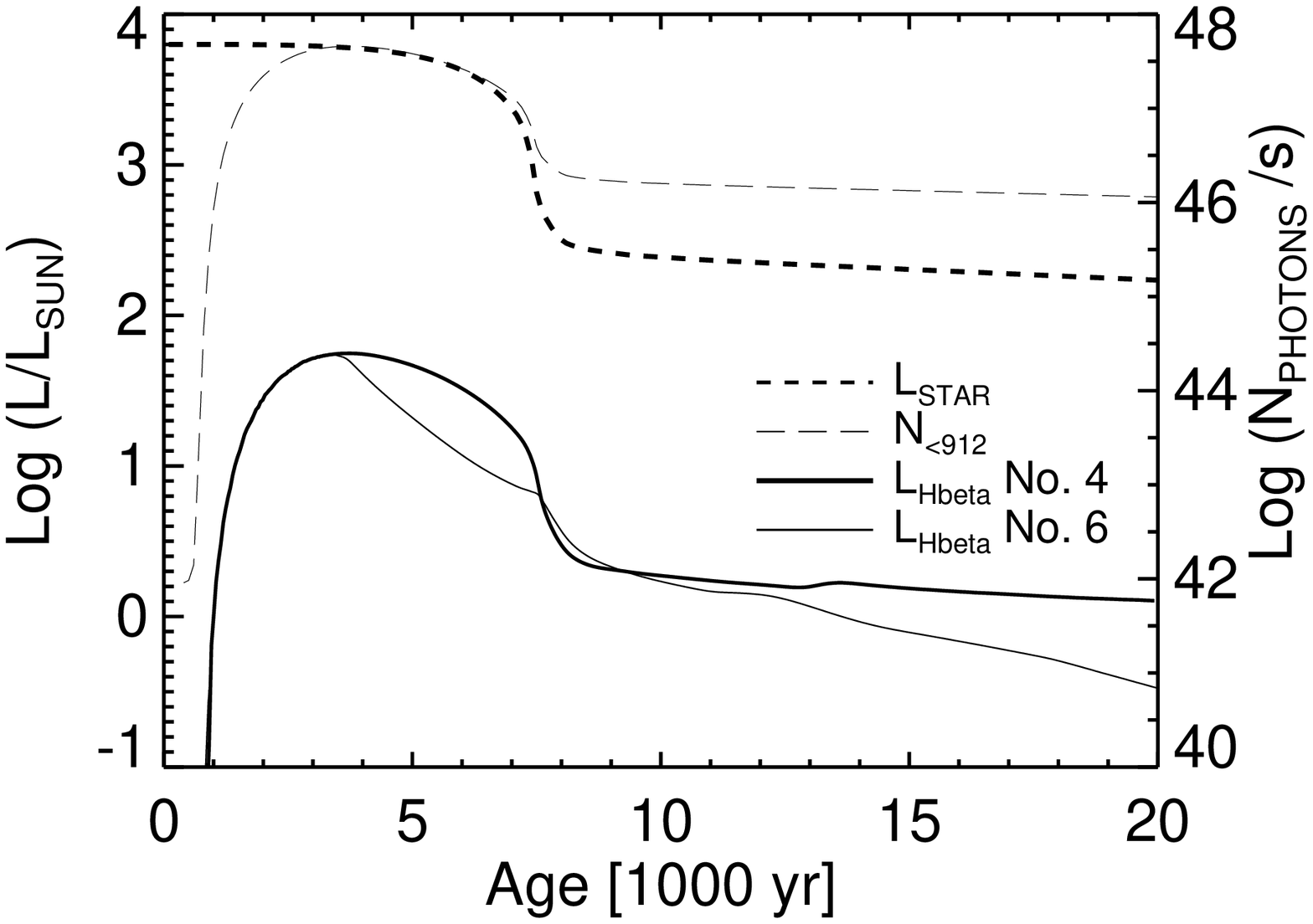}
\includegraphics[bb= 0.5cm 0.5cm 19.5cm 14.5cm, width=0.99\linewidth]{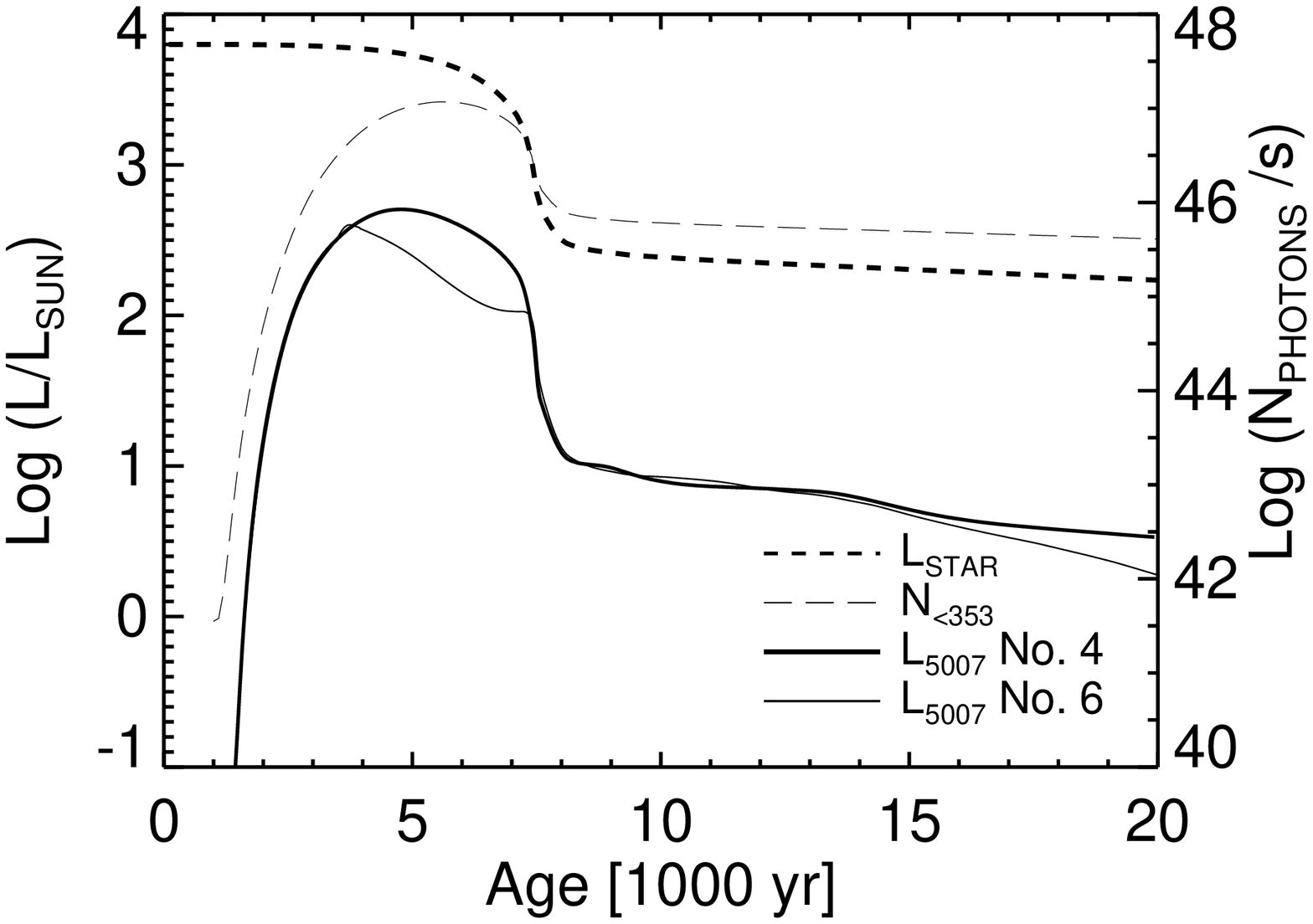}
\caption{Changes of stellar (thick dashed) and nebular line luminosities
	 (solid lines), $L$, and of the numbers of ionising photons
	 emitted per second, $N$ (long dashed), with post-AGB age
	 for a 0.605\,\Msun\ central-star model combined with two different envelope
	 models according to the sequences No.~4 (thick line) which never becomes
	 optically thin and No.~6 (thin line) which becomes thin to
	 ionising photons during the course of evolution.
         \emph{Top}:  evolution of $L_{\star}$, $L({\rm H}\beta)$, and $N_{<912}$
	 (right ordinate).
	 \emph{Bottom}: the same but for $L_{\star}$, $L(5007)$,
	 and $N_{<353}$ (right ordinate).
                     }
\label{evol.lum}
\end{figure}

  The influence of the optical depth is highlightened in
  Fig.~\ref{evol.lum} which shows the evolution of the nebular \hb\ and \oiii\
  line emissions with time for the 0.605~\Msun\ central star model coupled
  to two different nebular models (sequences Nos. 4 and 6).  In the top panel
  the number of photons emitted per second below 912~\AA, $N_{<912}$,
  is compared with the H$\beta$\ luminosity, whilst in the bottom panel the
  relevant quantities for \oiii\ are given: $N_{<353}$ and $L(5007)$, respectively.
  For comparison, the stellar luminosity is also displayed in both panels.

  For the optically-thick case of sequence No.~4, the line luminosities follow closely
  the number of ionising photons emitted per second, as expected from theory.
  While for hydrogen the maxima of the \hb\ emission and $N_{<912}$ occur at the
  same time (and at the same stellar temperature), the \oiii\ line emission peaks
  \emph{before} $N_{<353}$.
  The reason is that, at the high stellar temperatures involved,
  the prevailing ionisation stage shifts slowly from O$^{+2}$ to O$^{+3}$
  (see Sect.~\ref{oiii} for a more thorough discussion).

  As long as the models of sequence No.~6 are optically thick for ionising photons,
  their line emission is almost identical to that of sequence No.~4.  When the models
  become thin, their line emission drops below the value of an corresponding optically thick
  model: the ionised masses become smaller,
  and the nebula is said to be `density bounded'.
  Later on, after about 7\,000 years of post-AGB evolution,
  the ionisation drops in line with the decreasing central-star luminosity,
  and the outer parts of the models recombine to a large extent.  Consequently the
  nebular shell becomes able to absorb nearly all ionising photons again, and both
  sequences show virtually the same line luminosity for some time.

  The small differences between both sequences, after recombination
  has started, have several causes.   Firstly, recombination is not
  complete, i.e. the outer regions of the models of sequence No.\,6 do not become
  fully neutral (in hydrogen).  Secondly, we have to consider
  non-equilibrium effects due to the rapid fading of the stellar luminosity.
  Because the two sequences shown in Fig.~\ref{evol.lum} have different densities,
  recombination occurs with different time scales.  In particular, the models of
  sequence No.~6 are less dense than the corresponding ones of sequence No.~4, and
  consequently they recombine more slowly.
  Thirdly, because of their expansion the models of sequence No.\,6
  begin to reionise, reducing thereby their optical depth to such an extent that
  their line emission falls again below that of sequence No.~4
  (cf. $L({\rm H}\beta$) for $t\ga 10\,000$~yr).  At this particular evolutionary
  phase where the stellar luminosity and temperature changes only
  very little, the degree of ionisation is exclusively controlled by expansion.

  Oxygen recombines faster than hydrogen, thus its ionisation state can more easily
  keep pace with rapidly changing ionisation rates.  Indeed, after recombination
  the \oiii\ luminosities of both sequences are virtually the same
  (see Fig.\,\ref{evol.lum}, bottom).  But beyond \hbox{$t\approx 15\,000\,{\rm yr}$}
  the models of sequence No.~6 become optically thin again by re-ionisation and
  their \oiii\ emission falls below that of sequence No.~4.

  Since one has to consider that most observed PNe are optically thin to different
  degrees, one has to take into account the thick/thin transition for any
  realistic modelling of the PNLF.  Using our hydrodynamical simulations we will
  discuss in Sects. \ref{hbeta} to \ref{oiii} in more detail how the line emission of
  hydrogen, helium, and oxygen depends on the different model properties.

\subsection{The \hb\ luminosity}                             \label{hbeta}
  We begin with a discussion of the \hb\ luminosity displayed in
  Fig.~\ref{lhbeta} for four different model sequences selected from
  Table~\ref{tab.mod}.  The models of the sequences Nos.~4 and 10 remain always
  optically thick while those of sequences Nos.~6 and 6a, both with the
  same \textsc{Type C} initial envelope,
  become optically thin for Lyman continuum photons at stellar effective
  temperatures between 45\,000 and 60\,000~K.

\begin{figure}[!t]          
\includegraphics[bb= 2.0cm 4.5cm 19.5cm 19.2cm,  angle=-90, width=\linewidth]{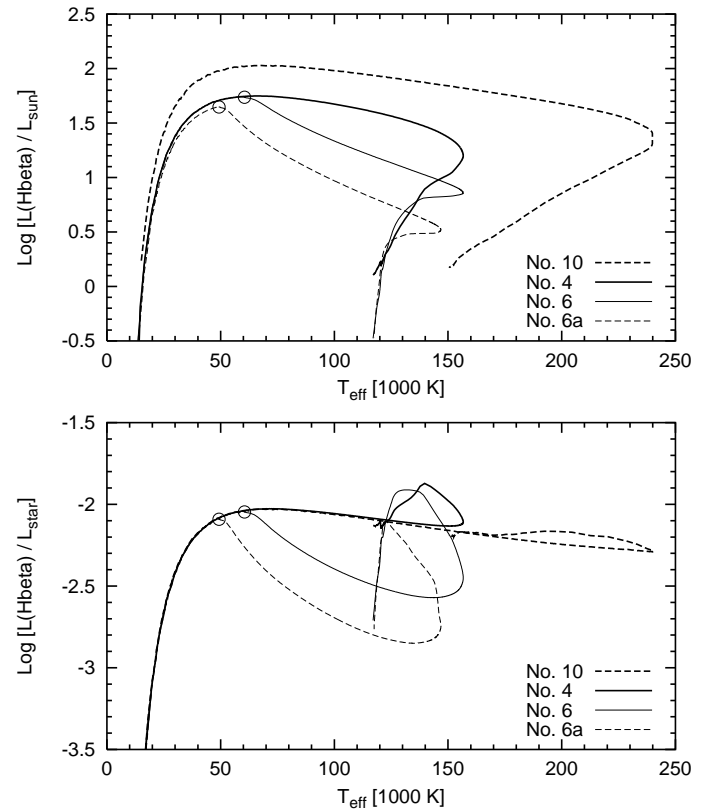}                  
\caption{$L({\rm H}\beta)/\lsun$ (\emph{top}) and $L({\rm H}\beta)/L_\star$
         (\emph{bottom}) vs.\ stellar effective temperatures for four model
	 sequences from Table~\ref{tab.mod} with 0.696~\Msun\ (No.~10),
	 0.605~\Msun\ (No.~4), 0.605~\Msun\ (No.~6), and 0.595~\Msun\ (No.~6a).
         The circles mark the transition from the optically thick to the
	 optically thin stage (Nos. 6 and 6a only, see text for details).
	 The endpoints of the tracks correspond to post-AGB ages of 20\,000 years
	 (Nos. 4, 6, 6a) and 3\,000 years (No.~10), respectively.
		     }
\label{lhbeta}
\end{figure}

  It is known from photoionisation models that the transition from the optically
  thick to the thin stage is not well defined.  The ionisation front is not sharp
  in a mathematical sense, and the model's optical depth depends also on the
  wavelength considered.  Even if a model is still optically thick, high-energy
  photons with their lower absorption probability may well escape.  A
  rather illuminating discussion of the optical depth problem in gaseous nebulae
  has been presented by \citet{Gru.00}.  In the present work we define the optically
  thick/thin transition in a rather pragmatic way as the moment when the 
  \ion{H}{i/ii} ionisation front passes the leading shock.  This moment is marked
  by circles in Fig.~\ref{lhbeta}.

  Figure~\ref{lhbeta} (top panel) demonstrates that the two optically thick sequences
  (Nos.~4 and 10) reach maximum \hb\ luminosities at a stellar effective temperature
  of about 65\,000~K, exactly where also $N_{<912}$ peaks due to
  the evolution of the stellar radius with effective temperature as the central star
  evolves across the HR diagram (see Fig.~\ref{photon.605}).

  In the bottom panel of Fig.~\ref{lhbeta} the evolutionary influence of the
  central star is taken out, and one sees immediately that 
  the maximum efficiency of converting Lyman photons into  \hb\ photons occurs at a
  stellar temperature of about 70\,000~K.
   At fixed stellar luminosity, $N_{<912}$ reaches a maximum at about 70\,000~K because
  of the following reason:  In general, the total number
  of stellar photons emitted per second, $N_{\rm total}$, decreases with increasing
  effective temperature because the fraction of high-energy photons becomes larger.
  Now, at lower temperatures, $N_{<912}$ increases very rapidly with
  temperature, and so does the \hb\ luminosity.  At very large temperatures, however,
  $N_{<912}$ approaches $N_{\rm total}$ and must finally decrease with temperature
  like $N_{\rm total}$.

  The existence of a maximum \hb\ luminosity (or flux) at intermediate
  stellar temperatures has already been shown, but not explained, by \citet{DJV.92}
  using a series of optically thick photoionisation models.
  The optimal efficiency of converting stellar luminosity into \hb\ line luminosity
  is about 1\,\% (bottom panel of Fig.~\ref{lhbeta}).  This value, of course, will
  not be reached if the PN shell becomes optically thin, as is the case for
  the sequences Nos.~6 and 6a.

  Once the central star has reached the hottest point of its evolution, it may
  shrink so rapidly towards white dwarf dimensions
  that the outer, less dense parts of the nebular shell may deviate from
  ionisation equilibrium for some time, as already mentioned in the previous section.
  For instance, once the star has passed its maximum effective temperature, it fades
  with time scales, $L_\star/\dot{L_\star}$, depending on its mass:
  $\simeq$\,400~yr for 0.595~\Msun, $\simeq$\,200~yr for
  0.605~\Msun, and only about 20~yr in the case of 0.696~\Msun.

  Non-equilibrium phases are best seen in the bottom panel of
  Fig.~\ref{lhbeta}.  One sees clearly that the \hb\ luminosity of \emph{all}
  models, albeit to different extents, do not follow the stellar luminosity decline
  after maximum stellar temperature.
  The extended `loops' indicate that the recombination rates considerably
  exceed the ionisation rates during the rapid stellar luminosity decline.
  They are very pronounced for the sequences Nos. 6 and 6a, but still visible
  for the sequences Nos. 4 and 10 despite of their rather dense, optically
  thick nebular shells.  Recombination exceeds ionisation for about 1500 years
  for the 0.605~\Msun\ case (No.~4), but only about 40~years for
  0.696~\Msun\ (No.~6).  Except for the 0.595~\Msun\ models, the \hb\ luminosity
  exceeds then even the limit given by equilibrium ionisation.
  The latter is finally reached again after the evolution is slowing down
  while the  central star is settling onto the white-dwarf sequence.

  Our models show that non-equilibrium ionisation of hydrogen may last for quite
  a substantial fraction of the total PN lifetime (up to 10\,\%).  Considering
  the fact that heavier ions recombine much faster than hydrogen, the conclusion
  is that  line strengths measured relative to that of hydrogen become
  \emph{smaller} during the recombination phase, leading consequently
  to erroneous results when standard plasma analysis methods are applied.

  During the non-equilibrium recombination phase,
  the efficiency of converting stellar photons into \hb\ photons
  temporarily exceed the maximum value typical for equilibrium conditions
  (Fig.~\ref{lhbeta}, bottom).  This may also hold for optically thin cases
  (cf. sequence No.~6).
  The absolute \hb\ luminosities, however, remain always well below the maximum
  gained earlier at intermediate stellar temperatures (top panel of Fig.~\ref{lhbeta}).
  We conclude thus that \emph{the maximum \hb\ luminosity of a PN is
  achieved either at $\teff \simeq 65\,000$~K for the optically thick case, or
  at lower temperatures if the nebular envelope becomes optically thin earlier.}

\begin{figure}[t]          
\includegraphics[bb= 2.0cm 4.5cm 19.5cm 19.2cm, angle=-90, width=0.96\linewidth]{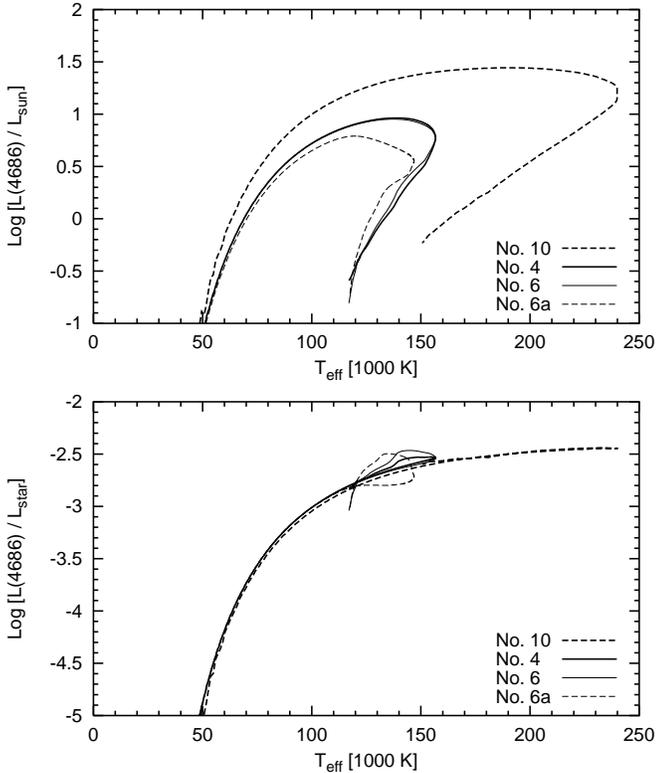}
\caption{$L({4686})/\lsun$ (\emph{top}) and $L({4686})/L_\star$
         (\emph{bottom}) vs.\ stellar effective temperatures for the same sequences
	 as in Fig.~\ref{lhbeta}.
}
\label{l4686}
\end{figure}

\subsection{The \mbox{\heii} luminosity}                     \label{heii}
  Although the helium lines are of no interest for luminosity functions,
  we present here, for completeness, the properties of the \heii\ 4686 \AA\ emission
  as they follow from our simulations (Fig.\,\ref{l4686}).
  By comparing with Fig.~\ref{lhbeta} we see that the \heii\ emission peaks at
  effective temperatures well beyond 100\,000~K, at a position that depends on
  the evolutionary properties of the central star, i.e. the larger its mass, the
  larger will be the peak temperature.    Only the 0.595\,\Msun\
  models (No.~6a) become optically thin for photons with
  $\lambda < 228$\,\AA\ once the temperature of the central star exceeds 120\,000~K.

  The maximum efficiency in converting stellar radiation into \heii\ 4686\,\AA\ line
  emission is only reached beyond 250\,000~K and is about 0.3\,\%
  (Fig.\,\ref{l4686}, bottom panel).  Because most central stars will not reach such
  high temperatures, and also because some
  nebulae will become optically thin for photons below 228~\AA\ (sequence No.~6a),
  the maximum absolute 4686~\AA\ emission is expected to occur at stellar temperatures
  between 100\,000 and 150\,000~K (top panel).

  Non-equilibrium effects are expected to be modest because He$^{+2}$
  recombines much faster than H$^+$.  Indeed, we see only small deviations from the
  equilibrium curves for the sequences Nos.~6, 6a, and also a very small `loop' for
  sequence No.~4, but not for the very dense 0.696~\Msun\ models of sequence
  No.\,10 (Fig.\,\ref{l4686}, bottom panel).

\begin{figure}[t]          
\includegraphics[bb= 1.9cm 4.5cm 19.5cm 19.2cm,  angle=-90, width=.95\linewidth]{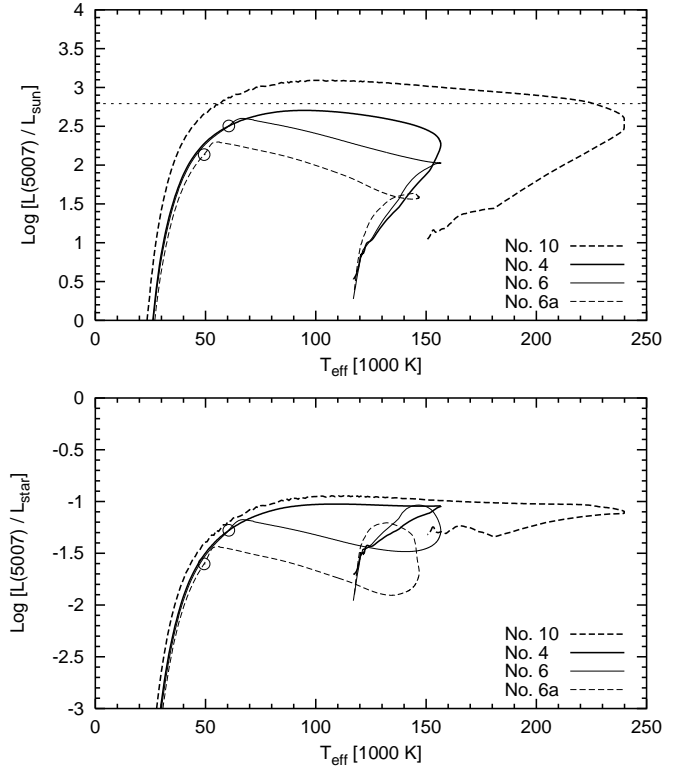}
\caption{$L(5007)/\lsun$ (\emph{top}) and $L(5007)/L_\star$
        (\emph{bottom}) vs.\ stellar effective temperatures for the same sequences
	as in Fig.~\ref{lhbeta}.
	The dotted horizontal line indicates $L(5007) = 620~\lsun$.
        The circles indicate where the models become optically thin for Lyman
	continuum photons (Nos. 6 and 6a only).
	}
\label{l5007}
\end{figure}

\subsection{The \protect{\oiii} 5007\,\AA\ luminosity}           \label{oiii}
  Understanding the behaviour of the \oiii\ 5007~\AA\ emission from the nebular shell
  as the central star evolves across the HR diagram is the key for any understanding
  of observed PNLFs.  The interpretation of this emission is more difficult since
  it depends strongly on the ionisation structure of oxygen which
  is a strong function of the evolutionary state  (see Figs.~\ref{ion1}--\ref{ion4}).

  The run of the \oiii\ 5007~\AA\ luminosity along  different evolutionary tracks
  is displayed in Fig.~\ref{l5007}.
  The absolute \oiii\ luminosity peaks already between 95\,000 and 100\,000~K (top panel),
  while the efficiency of converting stellar luminosity into 5007\,\AA\ line
  radiation reaches a maximum close to 110\,000~K (bottom panel), although
  $N_{<353}/N_{\rm total}$ peaks at about 180\,000~K, for the same reasons as
  discussed in Sect.~\ref{hbeta}.   The maximum conversion
  efficiency is about 10\,\% for the composition used here \citep[cf.][]{DJV.92},
  but does not change much until the stellar luminosity begins to fade.
  The more luminous 0.696~\Msun\ sequence shows a slightly larger conversion
  efficiency simply because of its  somewhat larger electron temperature.

  The position of the maximum of the conversion efficiency and the relatively
  flat run of the latter with stellar effective temperature is the result of
  two competing effects:  On one hand, the ionisation of oxygen shifts from
  O$^{+2}$ to O$^{+3}$ with increasing stellar temperature (see Figs.~\ref{ion2}
  and \ref{ion3}).  On the other hand,  the electron temperature rises steadily
  until the maximum stellar temperature is reached and compensates partly for
  the (relative) decrease of the emitting mass.  The electron temperature
  range is from 8\,000 to 12\,000~K for the 0.605~\Msun\ case (sequence No.~4) and
  from 9\,000 to 14\,000~K for the 0.696~\Msun\ case (sequence No.~10).

  The models of the sequences Nos.~6 and 6a become optically thin for
  O$^+$ ionising photons at 55\,000 and 65\,000~K, respectively, which means
  \emph{after} they became thin for Lyman continuum photons (circles).
  It is important to note that at this particular thick/thin transition the
  5007~\AA\ luminosity attains its (absolute) maximum (top panel of
  Fig.~\ref{l5007}).

  Because ionised oxygen recombines also much faster than H$^+$,
  we do not expect important non-equilibrium effects for the \oiii\ emission.
  Rather, the large `loops' apparent in Fig.~\ref{l5007} for the optically thin
  sequences are mainly the consequence of the changing ionisation structure of
  oxygen.  As already mentioned above, during the hottest, high-luminosity part of
  the stellar track the ionisation of oxygen is partly shifted towards the third
  stage (Fig.~\ref{ion3}).  When the stellar luminosity begins to drop, O$^{+2}$
  becomes temporarily
  the main ionisation stage, leading for a brief moment to a local maximum
  of the O$^{+2}$ mass and of $L(5007)/L_{\star}$ (cf.~Fig.~\ref{l5007}, bottom).
  With further decreasing stellar luminosity, recombination continues from
  O$^{+2}$ to O$^{+}$, and $L(5007)$ drops very rapidly.

  Due to the significant fraction of O$^{+}$ expected in nebular shells
  surrounding low-luminosity central stars (cf. Fig.~\ref{ion4}), the UV photon
  conversion efficiency drops considerably below its maximum value of $\approx 0.1$,
  as is seen for sequences Nos. 6 and 10 in Fig.~\ref{l5007} (bottom).
  This is different from the hydrogen case where the conversion efficiency is,
  in the optically thick case, only a function of the stellar effective temperature
  provided ionisation is in equilibrium (cf. Figs.~\ref{lhbeta} and \ref{l5007}).
  The \oiii\ luminosities of the models from the sequences Nos. 6 and 6a lie during
  their recombination `loops' temporarily somewhat \emph{above} the luminosities of
  the models from sequence No.~4, indicating a small imbalance between recombination
  and ionisation during the rapid luminosity decline of their central stars.

  The luminosity evolution of our hydrodynamical models as shown in the upper panel
  of Fig.~\ref{l5007} differs completely from the predictions of \citet{Ma.04}.
  According to their nebular models, which become optically thin very early,
  the 5007~\AA\ line luminosity \emph{increases}
  steadily during the crossing of the Hertzsprung-Russell diagram.
  Shortly \emph{before} the turning point is reached, the models become optically
  thick due to the reduced supply of ionising photons, and
  this thin/thick transition constitutes the moment of maximum \oiii\ line emission
  since afterwards the emission always decreases 
  for the same reasons as in the case of \hb\ \citep[see Fig.~10 in][]{Ma.04}:
  the stellar luminosity decreases, and the UV photon conversion efficiency is
  below its maximum as well.

  In conclusion we state that, for the metallicity chosen here, \emph{the
  maximum \oiii\ line emission of a PN occurs at $\teff\simeq 95\,000\dots100\,000$~K
  for optically thick configurations, or at lower temperatures if the PN becomes
  optically thin.}

\begin{figure}[t]            
\includegraphics[angle= -90,width= 1.03\linewidth]{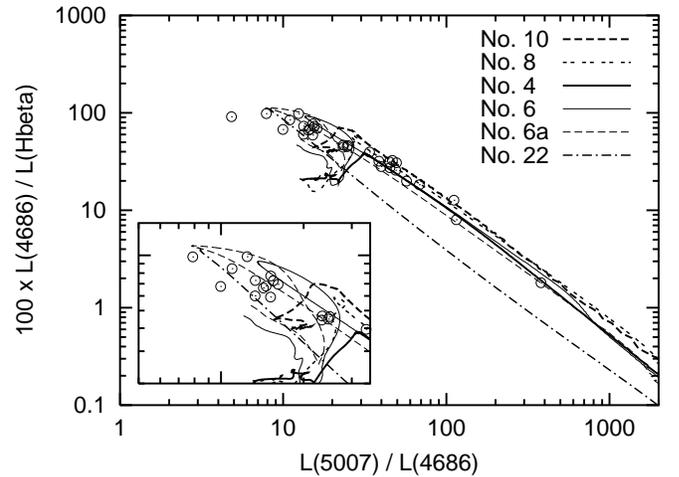}
\caption
  {{Correlations of line ratios between \hb, \heii\ 4686\,\AA, and \oiii\ 5007\,\AA,
  as predicted by our models and as observed for highly excited PNe (circles).
  The evolution proceeds from the right lower corner with low 4686/\hb\ towards
  large 4686/\hb\ values at maximum stellar temperatures.  The inset provides an
  enlarged view of the highest-excitation domain.
  The circles represent LMC planetaries selected from \citet{medo.91a,medo.91b}.}
  }
\label{lin.rat}
\end{figure}

\subsection{Line ratios \label{line.ratio}}
  {Before continuing we will discuss line ratios between important emission lines
  i.e. between \hb, \heii\ 4686\,\AA, and \oiii\ 5007\,\AA.
  The relative strengths of these lines
  reflect the ionisation structure of the nebula, determined mainly by the
  intensity and spectral energy distribution of the stellar radiation field
  and the optical depth of the nebular shell.  For instance, we see from the
  theoretical predictions in Fig.\,\ref{lin.rat} that for any given
  4686/\hb\ line ratio, the corresponding 5007/4686 ratio becomes
  smaller with decreasing central-star mass:
  For more slowly evolving central stars the nebular shell becomes more diluted with
  increased ionisation, leading to smaller 5007/4686 line ratios.

  We compare our models with a sample of LMC PNe taken from \citet{medo.91a,medo.91b}.
  This sample has the advantage that it is homogeneous and that the line fluxes
  are representative for the whole object.
  We see from Fig.\,\ref{lin.rat} that the anti-correlation between 4686/\hb\ and
  5007/4686 shown by the Magellanic Cloud PNe sample seems very well be explained by our
  hydrodynamic simulations within a central-star mass range of 0.59\ldots0.63\,\Msun.
  Note that the observed very low 5007/4686 values ($\simeq 10$) are only reached by
  those sequences with optically thin models (sequences Nos. 6, 6a, and 22).

  We caution, however, that the anti-correlation shown in Fig. \ref{lin.rat} is
  not sensitive to the overall structure of the models:  \cite{Ma.01, Ma.04}
  demonstrated that their models are also able to match the observed anti-correlation.

\subsection{Nebular excitation}
\label{excite}
  A very useful tool to classify the line emission of a PN and to determine its
  evolutionary stage is the excitation, expressed by suitable emission line
  ratios.  We follow here the definition introduced by \citet{DJV.92} and
  write for the excitation parameter $E$:
\begin{displaymath}
 E = 0.45\,\left[L(5007)/L({\rm H}\beta)\right], \ \ \ \     \hspace{1.9cm} 0.0<E<5.0\,,
\end{displaymath}
\begin{equation}
 E = 5.54\,\left[L(4686)/L({\rm H}\beta) + 0.78\right], \ \hspace{1.0cm} 5.0\le E < 12.0\,.
\end{equation}
  Using this definition, we computed $E$ for our
  model sequences and plotted the result against the absolute \hb\ and 5007~\AA\
  magnitudes (Fig.~\ref{excite.mag}).  A similar diagram, although based on pure
  photoionisation models, has been introduced and used by \citet{DM.90}
  to study the evolution of Magellanic Cloud PNe.

\begin{figure}[t]             
\includegraphics[bb= 2.1cm 2.1cm 19.5cm 15.1cm, angle=-90, width=0.98\linewidth]{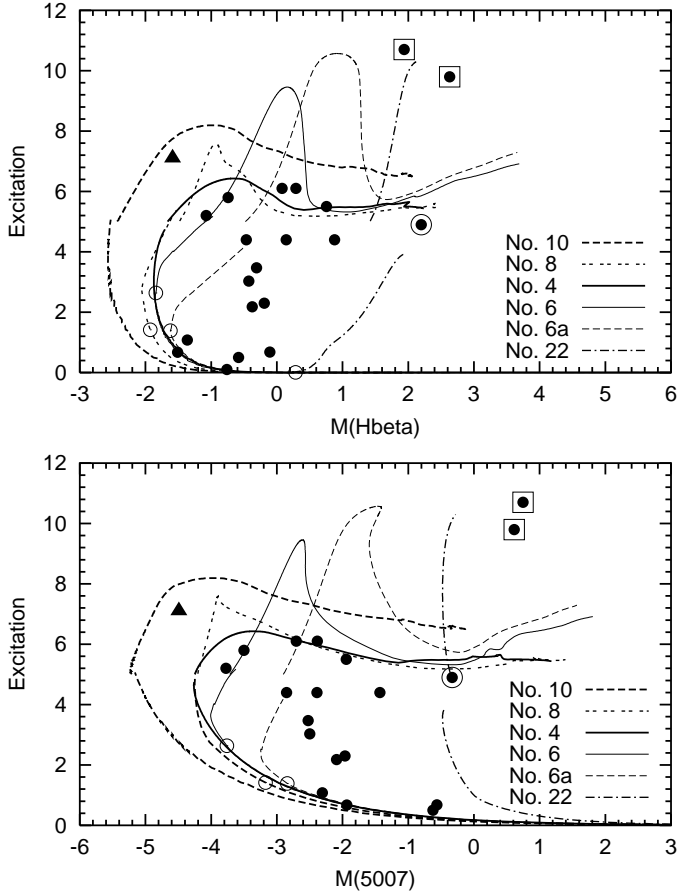}
\caption{Nebular excitation parameter, $E$, vs.\ absolute \hb\ (\emph{top}) and
         5007~\AA\ magnitudes (\emph{bottom}) for different hydrodynamical se\-quences
	 with 0.696~\Msun\ (No.\,10), 0.605~\Msun\ (Nos.\,4 and 6),
	 0.595~\Msun\ (No.\,6a), and 0.565~\Msun\ (No.\,22).  Absolute magnitudes are
	 based on Eq.~(1).  The small gaps between
	  $E\!\approx4$--$ 5$ for the sequences Nos.\ 6a and 22 are artifacts
	 due to the definition of $E$ (see text).  The open circles indicate again the
	 thick/thin transition of the nebular models (that of the 0.565~\Msun\
	 sequence is outside the plotted range).
	 Data of galactic PNe with spectroscopically determined distances are shown
	 for comparison \citep[][dots]{MKCJ.93}.   The `circled' dot belongs to
	 \object{NGC~7293}, the two `squared' dots to the high-excitation PNe
	 \object{NGC~1360} and \object{NGC~4361}.  The filled triangle marks
	 the position of \object{NGC~7027}.
        }
\label{excite.mag}
\end{figure}

  The evolution in Fig.~\ref{excite.mag} proceeds from low to high excitation, viz.
  the excitation increases generally with increasing stellar effective temperature.
  The maximum value of $E$ is reached at the highest possible
  stellar temperature because the $4686/{\rm H\beta}$ line ratio becomes
  largest there.   The actual value depends on the stellar mass, i.e. on
  the maximum achievable effective temperature, and on the optical depth of the
  nebular shell.  For example, the optically thin models of sequence No.~6 reach
  \mbox{$E\approx10$}, while the optically thick models of sequence No.~4 reach only
  \mbox{$E\approx7$}, despite the fact that both sequences are based on the same
  central-star model.  The reason for this difference is the fact that the
  optically thick models expand more slowly, remain much denser than the corresponding
  optically thin models and hence do not reach very high ionisation.
  During recombination, all sequences -- except No.~22 whose models do no recombine
  at all during our simulations because of their low density and their only slowly
  fading central star -- gain rather moderate excitations of
  $E\approx 6$ which later increase somewhat due to reionisation.

  Since we have seen in the previous sections that the \hb\ and
  5007\,\AA\ line luminosities peak at quite different effective temperatures
  (optical thick cases), the
  models predict also \emph{different} excitation parameters for these
  luminosity maxima.       
  For instance, the maximum line emission of \hb, occurring at
  \mbox{$\teff \simeq 65\,000$~K} for optically thick cases, corresponds to
  a rather modest excitation of $E\approx 3$--$4$ only.
  In contrast, the corresponding luminosity maximum
  for 5007\,\AA\ is reached at $E\simeq 5$ (\mbox{$\teff \simeq 100\,000$~K}).
  We see also from the models that the \oiii\ bright cutoff of
  $M^{\star}(5007) = -4.45$ corresponds to
  $M^{\star}({\rm H}\beta)\approx -2.2\dots-\!2.1$ (see also Fig.\,\ref{Hbeta.PNLF}).

  Figure~\ref{excite.mag} compares also the model predictions with observations.
  We used the sample of galactic PNe
  presented by \citet{MKCJ.93} for which individual distances based on detailed NLTE
  spectroscopic analyses of the stellar photospheres are available.  We computed
  the excitation parameter $E$ in the same way as we did for the models, using
  the published line ratios.
  We added \object{NGC~7027} since its distance is rather well known, too.

  Most of the objects are enclosed by sequence No.~6, i.e.\ by the
  0.605~\Msun\ post-AGB model with the \textsc{Type C} envelope, together with
  sequence No.~22 (0.565~\Msun), indicating a rather small mass range as one
  would expect.  Many objects have rather high nebular excitations ($E\ga3$)
  but only moderately hot central stars \mbox{($\teff \la 70\,000~{\rm K}$)}, which is
  only possible if their nebular shells are optically thin for Lyman continuum
  photons.  \object{NGC~7293} has a position consistent
  with the idea that it is in the late recombination or early reionisation stage.
  The two objects with \mbox{$E \ga 10$}, \object{NGC~1360} and \object{NGC~4361},
  can well be explained by a low-mass sequence similar to No.~22.   Its low-mass
  central star evolves so slowly that the nebular shell becomes very diluted
  and gains a very high excitation during the evolution across the
  Hertzsprung-Russell diagram.

  \object{NGC~7027} is an interesting object since it belongs to the class of PNe
  with a relatively massive central star and with an optically
  thick nebular shell.  Its position in Fig.~\ref{excite.mag} suggests a
  central-star mass close to 0.7~\Msun, in good agreement with the
  independent estimates of \citet{Latetal.00}.  By chance, its \oiii\ line emission
  ($M(5007)= -4.49$) is at the observed bright cut-off.
  Judging from our sequence No.\,10, \object{NGC~7027} must have been brighter
  in the past, with a maximum brightness of $M(5007) \approx -5$.  It can be
  estimated from our simulations that the total time spent between 
  $M(5007)\simeq -4.5 \dots -5$ is about 500~yr.

\begin{figure}[t]         
\includegraphics[bb= 2.1cm 2.1cm 19.5cm 15.1cm, angle=-90, width=0.96\linewidth]{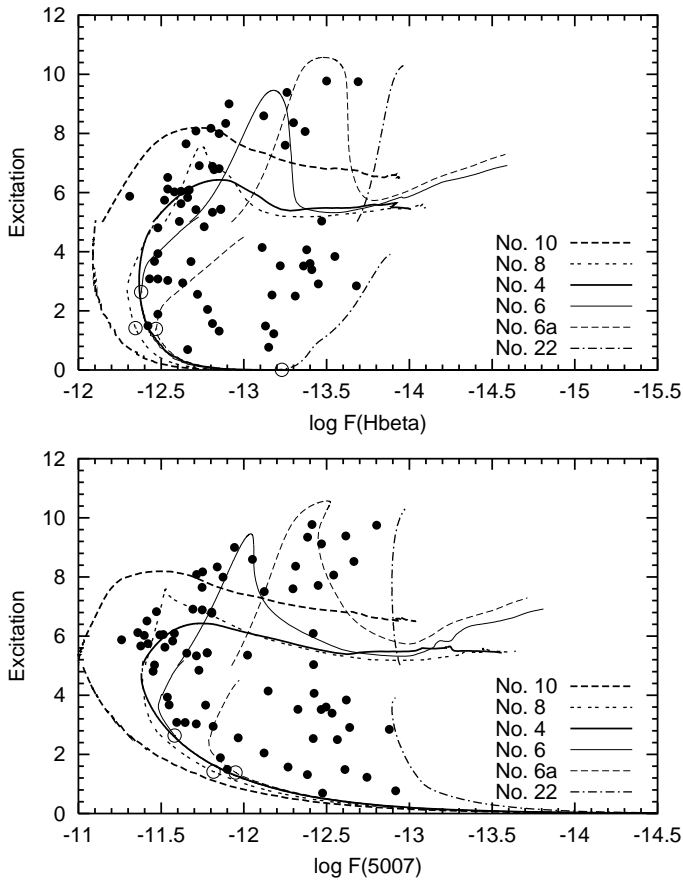}
\caption{Nebular excitation parameter, $E$, vs. line fluxes
        (in erg\,cm$^{-2}$\,s$^{-1}$) of H$\beta$\ (\emph{top}) and 5007\,\AA\
	(\emph{bottom}) for a sample of LMC planetaries (dots) taken from the lists
	of \citet{medo.91a,medo.91b}.  The evolutionary sequences are the same as in
	Fig.~\ref{excite.mag}.  We scaled the line fluxes of the models assuming
	a LMC distance of 47\,kpc and an extinction of $c=0.28$
	\citep{DJV.92}.
        }
\label{excite.LMC}
\end{figure}

  Of course, the best PNe sample for testing theory is provided by the
  Magellanic Clouds because the distances are known.  We selected a subsample from
  the Large Magellanic Cloud, consisting of objects from \cite{medo.91a, medo.91b}
  with accurate spectrophotometry, and determined the excitation parameters according
  to above prescription.  The result is shown in Fig.\,\ref{excite.LMC} where the
  excitations are plotted over the observed line fluxes of \hb\ (top) and 5007\,\AA\
  (bottom).
  The metallicity of LMC objects is only moderately subsolar, and any
  complications that may arise from metallicity differences between our models
  and the objects placed in Fig.\,\ref{excite.LMC} are expected to be insignificant.
  Figure~\ref{excite.LMC} is very similar to Fig.~8 of \citet{DM.90}, but with the
  important difference that we compare the observations with hydrodynamical
  models, allowing a more thourough interpretation.
  The hydrodynamical sequences are the same as in Fig.\,\ref{excite.mag}.

  First of all, Fig.\,\ref{excite.LMC} is fully consistent with Fig.\,\ref{excite.mag}.
  The LMC PNe are confined in the excitation--flux plane by models with
  central stars of about 0.63~\Msun\ on the high-mass side and by models with
  central stars not less massive than about 0.57~\Msun\ on the low-mass side.
  This finding is in contrast to \citet{DM.90} who claimed, based on their Fig.\,8,
  that high-excitation objects \mbox{($E\simeq 8\ldots10$)} with smaller line fluxes belong
  to optically thin nebulae around massive \mbox{($\simeq 0.7$\,\Msun)} central stars.
  However, \citeauthor{DM.90} did not consider that, once an optically thick object
  has gained its highest excitation (or largest stellar temperature),
  it will remain optically thick because of the stellar luminosity
  drop (cf.\ the sequences Nos.~4 and 10 in Fig.\,\ref{excite.LMC}).  The comparison
  between sequence No.~4 and No.~6, both with the same central-star model of
  0.605\,\Msun, reveals clearly that very highly excited objects with medium line
  fluxes must have optically thin nebulae with normal central stars of about 0.6~\Msun!

  There appears to be one exception, viz. \object{SMP 62} which is
  the brightest object in both \hb\ and \oiii\ 5007\,\AA\ while the excitation is only
  moderate, $E\simeq 6$.  This object could well be a similar high-mass object
  with an optically thick nebula like \object{NGC~7027}.

  For the interpretation of the luminosity function it is important to realize from
  Fig.\,\ref{excite.LMC} that the objects brightest either in H$\beta$\ or \oiii\ are
  \emph{not} those with the largest excitation parameters  \citep[see also][]{MKCJ.93}.
  Instead, the PNe brightest
  in H$\beta$\ have only $E\approx 2\ldots4$ (except \object{SMP 62}),
  while the objects brightest in 5007\,\AA\ have $E\simeq 6$,
  in excellent agreement with the predictions of our hydrodynamical simulations.

  The bright end of the Magellanic Cloud PNLF is obviously populated by
  PNe with central stars between ${M\simeq}\ 0.60\dots0.63$~\Msun\ with nebular shells
  which are optically thick, or partly thick, during the
  \emph{whole} evolution across the Hertzsprung-Russell diagram.  Nebulae around
  slightly less massive central stars become optically thin and fainter during their
  evolution, as indicated in Fig.\,\ref{excite.LMC} by our sequences Nos.\,6 and 6a.

  The \citet{Ma.04} models are expected to behave differently in an
  excitation--flux diagram. As already mentioned earlier, their nebular models are
  optically thin for Lyman continuum photons already at very low effective
  temperatures even for large central-star masses.  Their maximum \oiii\ emission
  occurs when recombination sets in close to maximum stellar temperatures where
  the excitation is expected to be large ($E \ga7$).
  Thus the \citeauthor{Ma.04} models predict maximum line fluxes, or maximum
  magnitudes, at these high excitation levels, which is not observed.
  Since their \oiii\ line luminosity beyond stellar temperatures of 100\,000~K
  falls off, central-star masses up to 0.75\,\Msun\
  are needed to provide the observed 5007\,\AA\ cut-off brightness.

  We conclude again, based on the excitation--flux/magnitude distributions shown in
  Figs.\,\ref{excite.mag} and \ref{excite.LMC},
  that {the \citeauthor{Ma.04} models do not provide an
  adequate description of the observed properties of planetary nebulae.}

  A more direct comparison between theory and observation is possible, too.
   \citet{DM.91a, DM.91b} derived by means of detailed photoionisation modelling
   that the \oiii\ brightest objects in their LMC sample ($E\simeq 5\ldots6$) have
   high-luminosities ($L\simeq5000\ldots8500$ \Lsun) and hot
   ($\teff\simeq 100\,000\ldots130\,000$ K) nuclei.  This is exactly what our
   sequences Nos. 4 and 8 predict (Fig.~\ref{excite.LMC}).  Some of these objects are
   also contained in the sample investigated by \citet{SSVM.06}, viz. \object{SMP 62,
   SMP73, SMP 74, SMP 88}, and \object{SMP 92}.  From the images and spectra one
   sees immediately (i) that these objects are quite extended, similar to the
   Galactic counterparts shown in Fig.~\ref{rel.rad}, and (ii) that they have rather
   weak \nii\ emission.   The latter fact indicates, in conjunction with the high
   luminosities, that these objects are not yet recombining.  This is again in
   contrast to the prediction of the \cite{Ma.01, Ma.04} models which achieve their
   maximum \oiii\ emission while they are recombining during the luminosity
   drop of their central stars \citep[see][Figs. 10 and 11 therein]{Ma.04}.

\begin{figure}[!t]          
\includegraphics[bb= 2.1cm 2.1cm 19.5cm 15.3cm, angle=-90, width=0.97\linewidth]{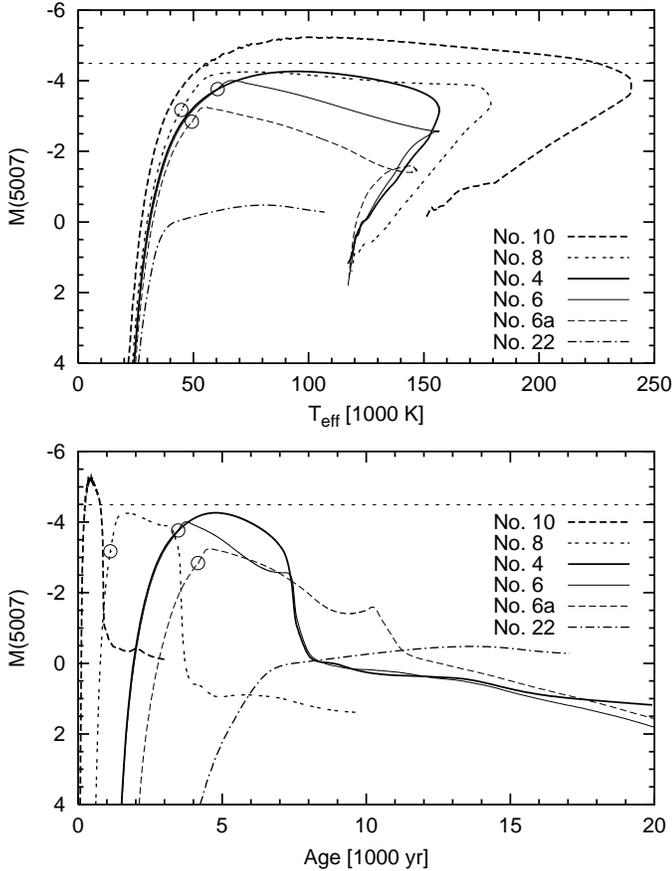}
\caption{Brightness evolution of the 5007\,\AA\ line for the sequences listed in
         Table~\ref{tab.mod} vs.\ stellar effective temperature (\emph{top})
	 and post-AGB age (\emph{bottom}).   The circles indicate the
	 moments when the model nebulae become optically thin in the hydrogen
	 Lyman continuum.  Note that the sequences No.\ 4 and 10 remain
	 optically thick during their whole computed evolution, while the
	 nebular models of sequence No.\ 22 turn into the optically thin stage
	 very early (outside the plotted range).
	 The horizontal dotted line indicates in both panels
	 the observed bright cut-off.
        }
\label{M5007}
\end{figure}

\section{The \oiii\ brightness evolution}                            \label{LF}
  For a better understanding of the general properties of the PNLF
  it is useful to look into the \oiii\ 5007\,\AA\ brightness evolution of our
  hydrodynamical sequences.
  Therefore, we plotted in Fig.\,\ref{M5007} $M(5007)$ against stellar temperatures
  (top) and post-AGB ages (bottom) for all sequences listed in Table~\ref{tab.mod}.
  The general behaviour is, independent of
  the total time span plotted, that they become bright very rapidly but fade more
  slowly.  The optically thin models (sequences No.~6 and 6a) show temporary
  brightness peaks when O$^{+3}$ recombines to O$^{+2}$ and then to O$^{+}$
  while the central star fades at the high-temperature end of its evolution.
  Finally, all models settle at brightnesses between 0 and 2 mag as the central star
  approaches the luminous end of the white-dwarf domain.

  The top panel of Fig.\,\ref{M5007} is the same as the top panel
  of Fig.\,\ref{l5007} but contains additionally the sequences No.~8 and 22
  which deserve an extra comment.  The nebular models around the
  0.625~\Msun\ central star (sequence No.\,8) become optically thin during the evolution
  across the Hertzsprung-Russell diagram, but the central star evolves so quickly
  that recombination sets already in before the whole computational domain can be
  ionised  \citep[cf. Fig. 17 in][]{Pe.04}.  Because of its optical depth being
  slightly below unity the models do not reach the cut-off brightness
  (horizontal dotted line), but remain brighter than
  $-4$~mag for about 1500~yr (Fig.\,\ref{M5007}, bottom).  On the opposite side,
  the models of sequence No.~22 with their very slowly evolving central star of
  0.565~\Msun\ never become brighter than $-0.5$~mag because they are optically
  very thin already in a very early phase of their evolution.

  The predictions from our hydrodynamical models concerning the expected brightness
  evolution can be summarized as follows:
\begin{itemize}
\item  Nebular shells around more massive central stars remain optically thick, or
       nearly thick, and their \oiii\ line emission is close to or even exceeding the
       observed cut-off brightness $\simeq -4.5$\,mag.
\item  PNe with less massive central stars will become optically thin for
       Lyman continuum photons, and their maximum \oiii\ line emission
       occurs shortly after this thick/thin transition.  Even if
       recombination sets in later,
       the \oiii\ brightness will \emph{not} again attain its previous maximum.
       The reason is the lower stellar luminosity in combination with the reduced
       conversion efficiency of stellar UV photons as discussed earlier.
\end{itemize}

  The bottom panel of Fig.\,\ref{M5007} has to be compared with Fig.\,10 of
  \citet{Ma.04}.  The differences are evident:  the \citeauthor{Ma.04} models
  are optically thin with comparatively low \oiii\ emission until the latter peaks
  during recombination as the central star begins to fade.  The peak values for
  nuclei of about 0.6~\Msun\ correspond roughly to the `recombination peaks' seen
  in the bottom panel of our Fig.\,\ref{M5007} (sequences Nos. 6 and 6a).
  It is thus evident from our figure that, in order to make these \oiii\ peaks
  as bright as the observed cut-off,
  relatively massive and luminous central stars with at least  0.7\,\Msun\ are needed.

\begin{figure}[!t]          
\includegraphics[bb= 1cm 0.8cm 19.0cm 14cm, width=\linewidth]{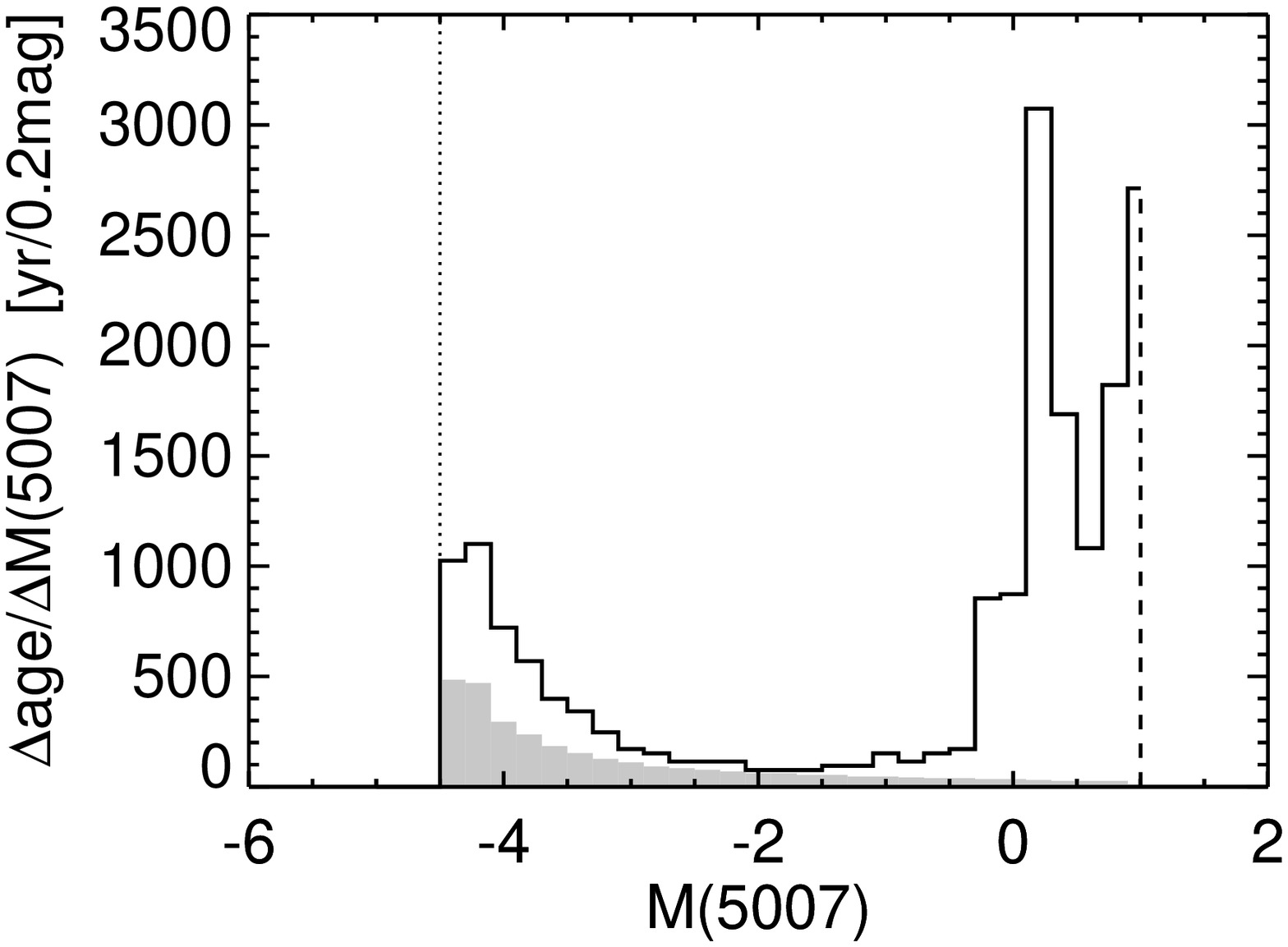}
\includegraphics[bb= 1cm 0.8cm 19.0cm 14cm, width=\linewidth]{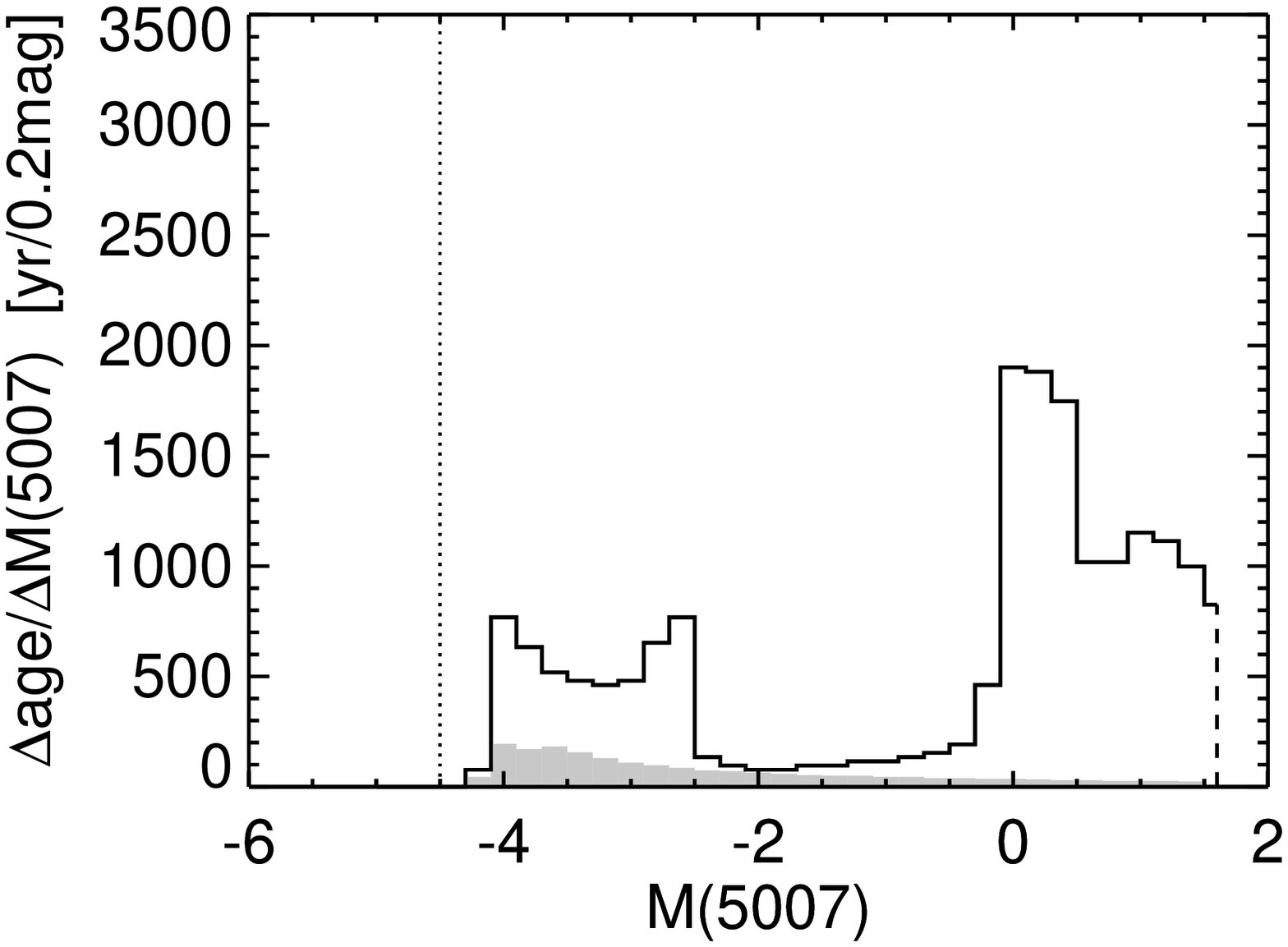}
\caption{Histograms of two individual \oiii\ 5007\,\AA\ luminosity function proxies,
         $\Delta age/\Delta M(5007)$, computed
         for a central star of 0.605~\Msun\ with two different initial circumstellar
	 envelopes as presented by sequences Nos.\,4 (\emph{top}) and 6 (\emph{bottom}).
	 Plotted are the total times spent per  absolute magnitude bin of
	 0.20~mag.   The gray areas reperesent the contributions
	 from increasing brightness.
	 The observed bright $M(5007)$ cut-off is indicated by the vertical dotted
	 line.  The vertical dashed lines at the faint end, $M(5007)\ge 0.5$,
	 indicates that there the distributions are affected by the limited simulation
	 times (cf. Fig.\,\ref{M5007}, bottom panel).
        }
\label{605.LF}
\end{figure}

\subsection{Individual contributions to the luminosity function}
\label{ind.LF}
  With our very limited number of detailed evolutionary sequences at hand we are
  not able to construct a theoretical luminosity function useful for interpreting
  observations.  Our computations are, however, very helpful in elucidating how
  different combinations of central stars and nebular shells will influence
  the shape of the luminosity function.
  To this end, we converted the `evolutionary' tracks from Fig.\,\ref{M5007}
  (bottom panel) into histograms giving the expected lifetime of a model per
  specified magnitude interval -- this presentation being a proxy for the
  (individual) luminosity function of each sequence.
  In the following figures we preferred a linear ordinate as to facilitate the
  visualisation of the correspondences with the bottom panel of Fig.\,\ref{M5007}.
  The usual logarithmic presentation can be found below.

  At first we illustrate
  in Fig.\,\ref{605.LF} the influence of two different nebular configurations
  coupled the the same stellar model of 0.605 \Msun.   The different shapes of the
  luminosity function seen in the figure are then entirely due to the different
  development of the optical depths, which in turn is the consequence of the different
  initial configurations with respect of density distribution and total mass
  (cf. Sect.~\ref{hydro.mod}).

   The optically thick case (top panel) is characterised
  by an U-shaped curve reflecting the evolution of the whole system:  one maximum
  corresponds to the bright turning point close to $M(5007)\simeq -4.5$ mag, the other
  at the faint end is due to the strongly reduced evolutionary speed when the
  central star enters the white-dwarf `cooling' path.  In the optically thin case
  (bottom panel) the luminosity function has a flatter shape at the bright end,
  combined with a `recombination peak' at $\simeq -2.7$ mag.  At the faint end,
  after recombination, the luminosity evolution is modified by re-ionisation.
  (cf. Fig.\,\ref{M5007}, bottom panel).  This faint part of the luminosity
  function, however, is of no observational interest.

  In the optically thick case (top panel) the evolution
  towards maximum \oiii\ line emission contributes about 50\,\% to the total
  luminosity function except around ${M(5007)\simeq -2}$ mag where this fraction is
  close to 100\,\%.  In the optically thin case (bottom), the evolution towards
  maximum line emission contributes much less to the total luminosity function.
  Both models populate
  the region of the bright cut-off, \mbox{$M(5007)= -4.0\ldots -4.5$ mag}, but for quite
  different periods:  2900~yr in the optically thick case (top, No.~4)
  but only about 800\,yr for the sequence No.~6 (bottom) whose models spend, in turn,
  much more time at medium brightnesses between \mbox{$M(5007)\simeq-3\dots-4$}.
  The deep depression between $M(5007)\simeq-2.5$ and 0 mag common for both
  luminosity functions is produced by the rapid luminosity drop of the central star
  (cf. also Fig.\,\ref{M5007}).

\begin{figure}[t]          
\mbox{
\includegraphics[bb= 1cm 0.8cm 19.0cm 14cm, width= 0.49\linewidth]{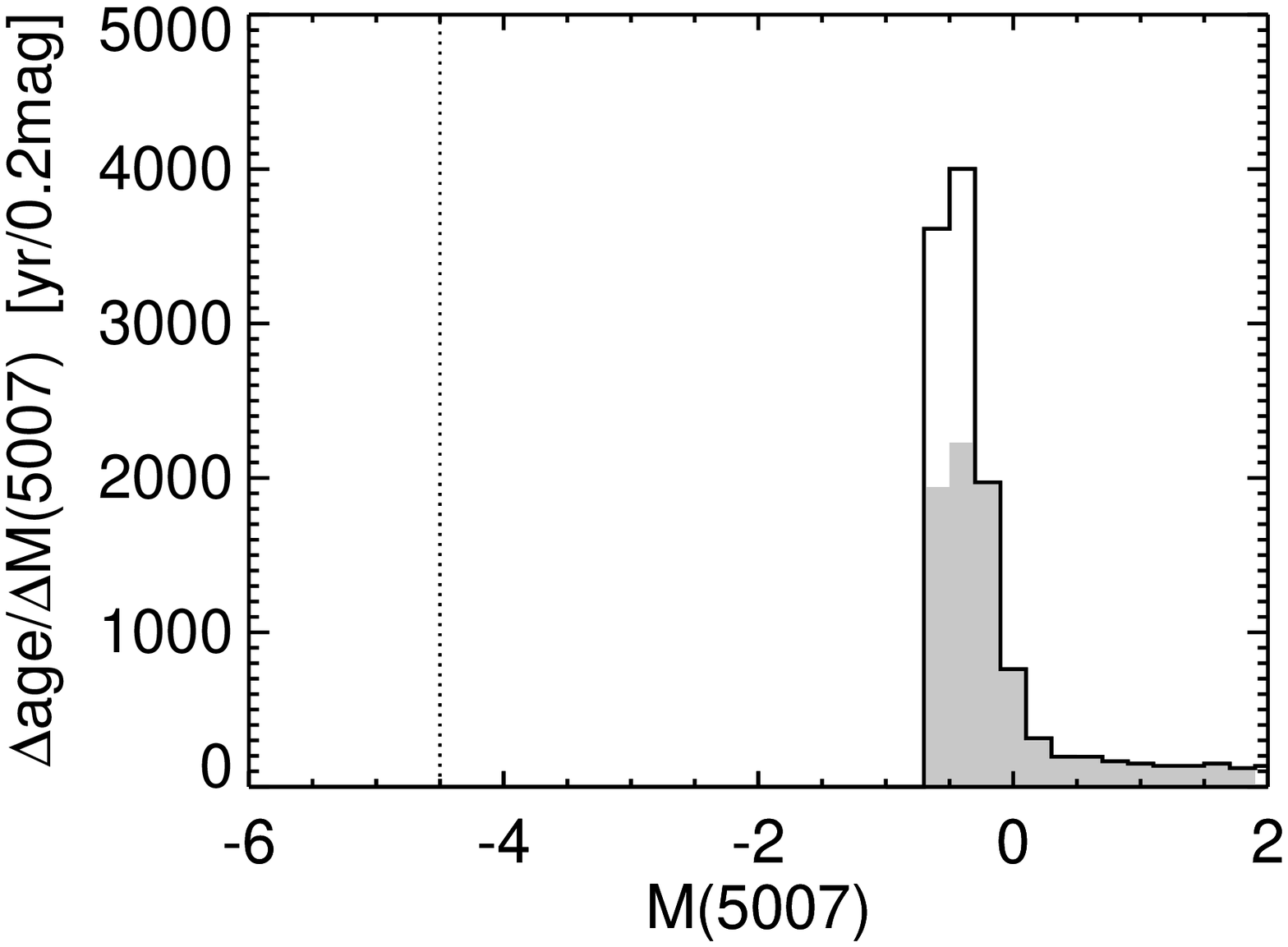}
\includegraphics[bb= 1cm 0.8cm 19.0cm 14cm, width= 0.49\linewidth]{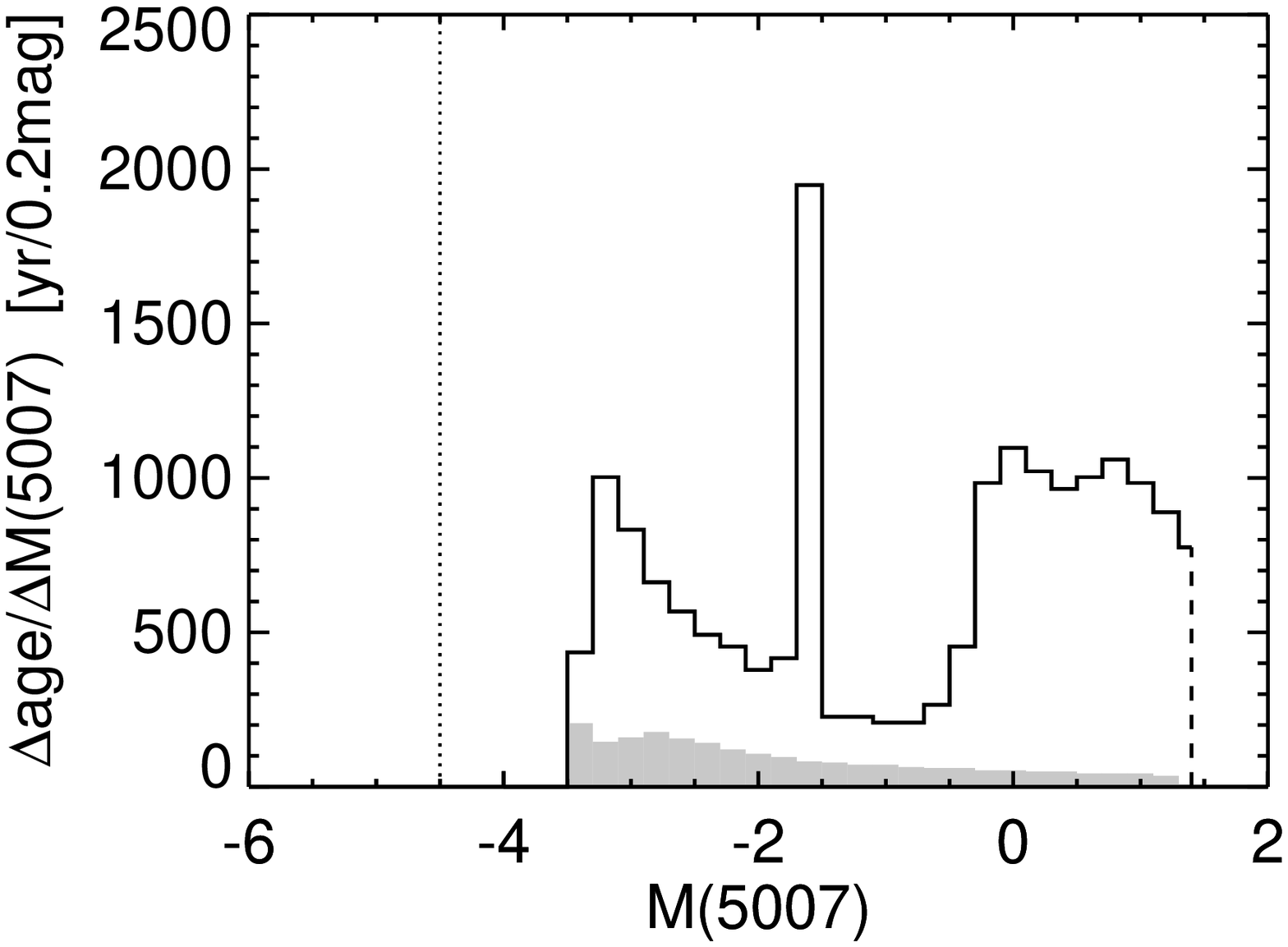}
     }
\mbox{
\includegraphics[bb= 1cm 0.8cm 19.0cm 14cm, width= 0.49\linewidth]{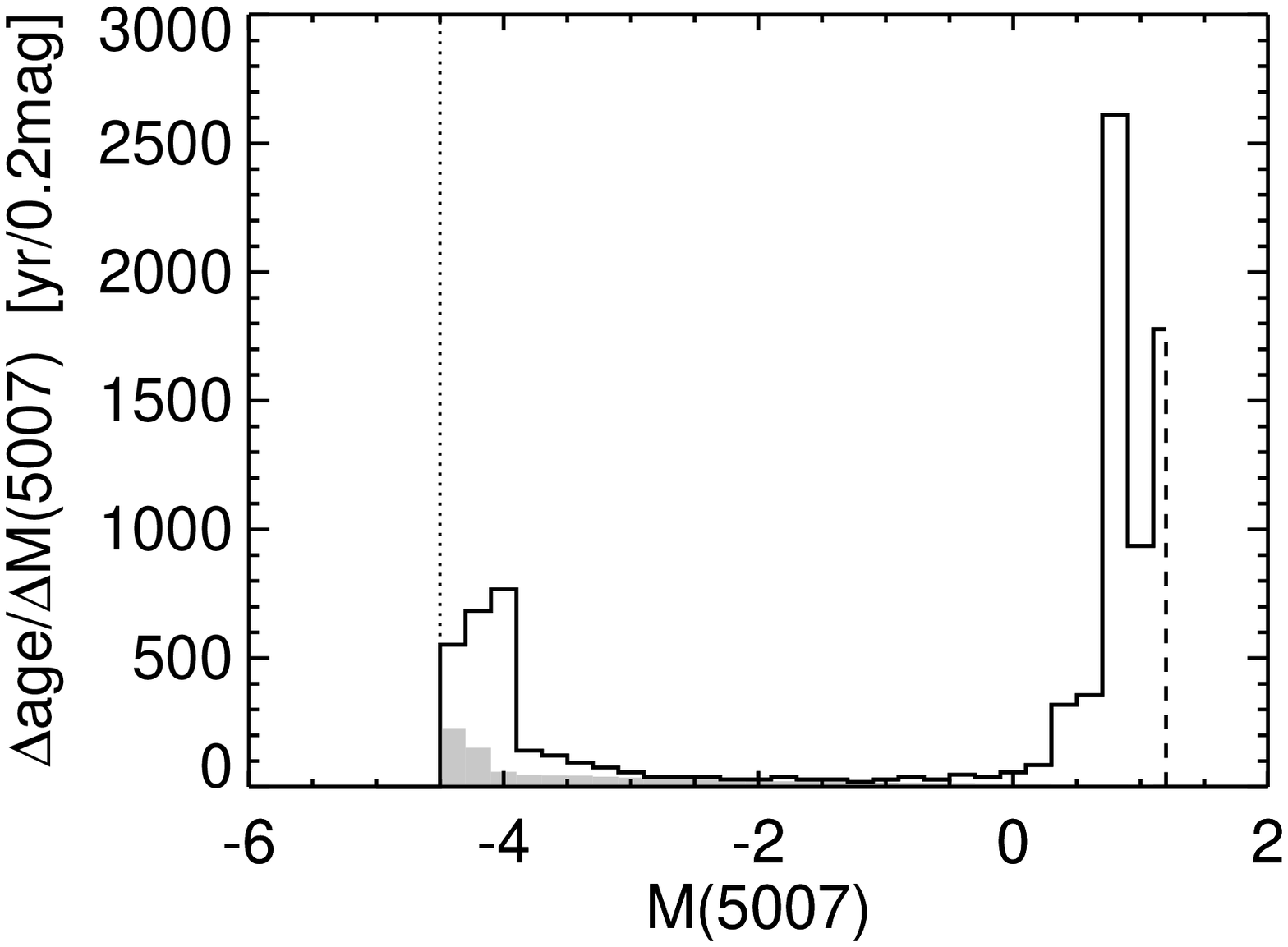}
\includegraphics[bb= 1cm 0.8cm 19.0cm 14cm, width= 0.49\linewidth]{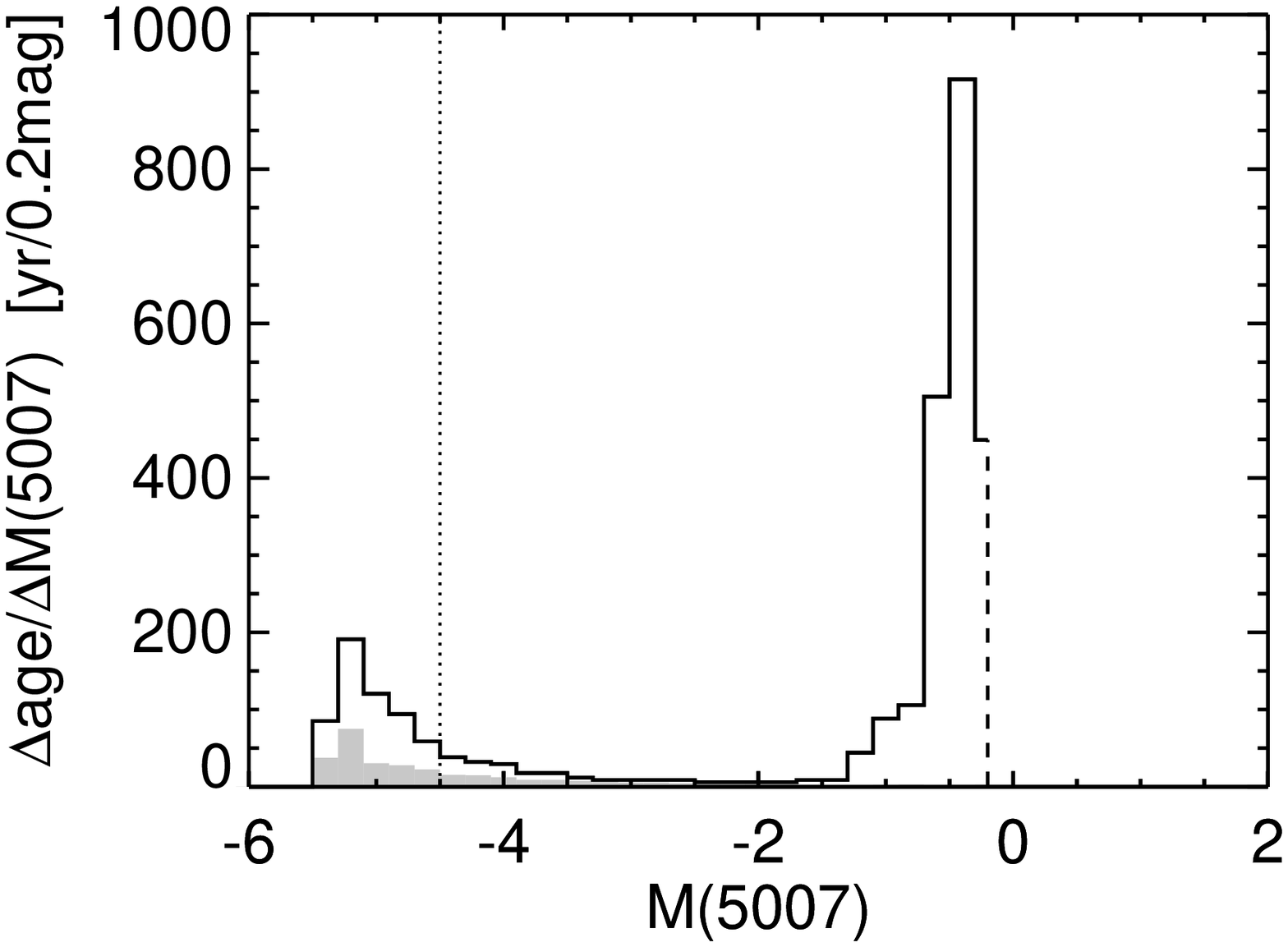}
     }
\caption{The same as in Fig.~\ref{605.LF}, but for 0.565~\Msun, (sequence No.~22,
         \emph{top left}), 0.595~\Msun, (sequence No.~6a, \emph{top right}),
	 0.625~\Msun, (sequence No.~8, \emph{bottom left}), and 0.696~\Msun,
	 (sequence No.\,10, \emph{bottom right}).  Note the different ordinate
	 ranges!
        }
\label{rest.LF}
\end{figure}

  The luminosity functions (resp. their proxies) of the other sequences from
  Table~\ref{tab.mod} are shown in Fig.\,\ref{rest.LF}.
  This figure demonstrates how sensitively the bright
  end of each individual luminosity function depends on the mass of the central star,
  viz. the speed of its evolution, and the time variation of the optical depth of the
  nebular shell, determined by gas density and expansion rate.
  Low-mass objects with, e.g. 0.565~\Msun, and optically very thin nebular envelopes,
  are contributing only a narrow but rather faint magnitude interval (top left panel).
  A typical object with a central star close to 0.6~\Msun\ becomes much brighter in
  $M(5007)$, but fails to reach the observed cut-off because the shell becomes
  optically thin.  Recombination peaks are in no case bright enough
  to contribute to the bright cut-off at $-4.5$ mag.
  Shells around more massive, i.e. faster evolving central stars, remain mainly
  optically thick, and their \oiii\ emission is able
  to reach or to pass the observed $M(5007)$ cut-off, even if the envelope becomes
  somewhat optically thin, as is the case for the 0.625~\Msun\ sequence (bottom left
  and right).

  However, the models with more massive central stars stay brighter than $-4.5$ mag
  in \oiii\ for only a brief period, viz.\ $\approx 500$~yr for 0.7~\Msun.
  The border-line models with a 0.625\,\Msun\ central star remain for about 1500~yr
  close to $M(5007) \simeq -4.5$ mag.
  Although PNe with such massive central stars do exist (e.g. \object{NGC~7027}),
  their significance for the bright end of the PNLF can only be determined by
  detailed population syntheses.
  In any case, our models predict much longer lifetimes at maximum \oiii\
  emission than the \citet{Ma.01} models.

  Recently \citet{CSFJ.05} estimated a time span of $\approx\! 500$~yr as
  a typical period to be spent by PN in the top 0.5 mag of the luminosity function.
  According to our simulations this time should be, on the average, considerably
  longer, with the consequence of a corresponding reduction of the fraction of
  stars with appropriate masses that have to evolve off the main sequence and that
  will eventually evolve into PNe populating the bright end of the luminosity function.

\begin{figure}[t]            
\mbox{
\includegraphics[width=0.5\columnwidth]{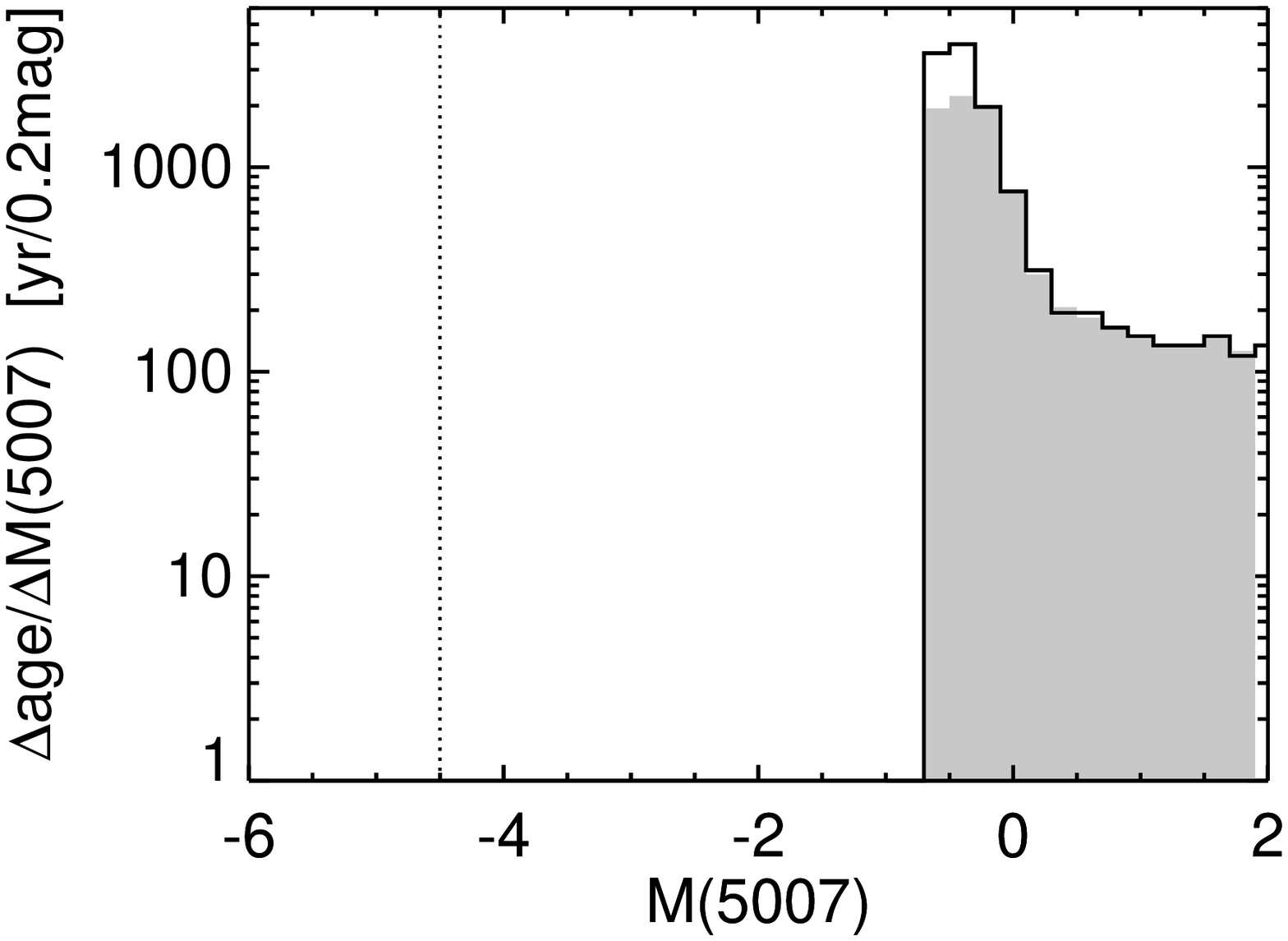}
\includegraphics[width=0.5\columnwidth]{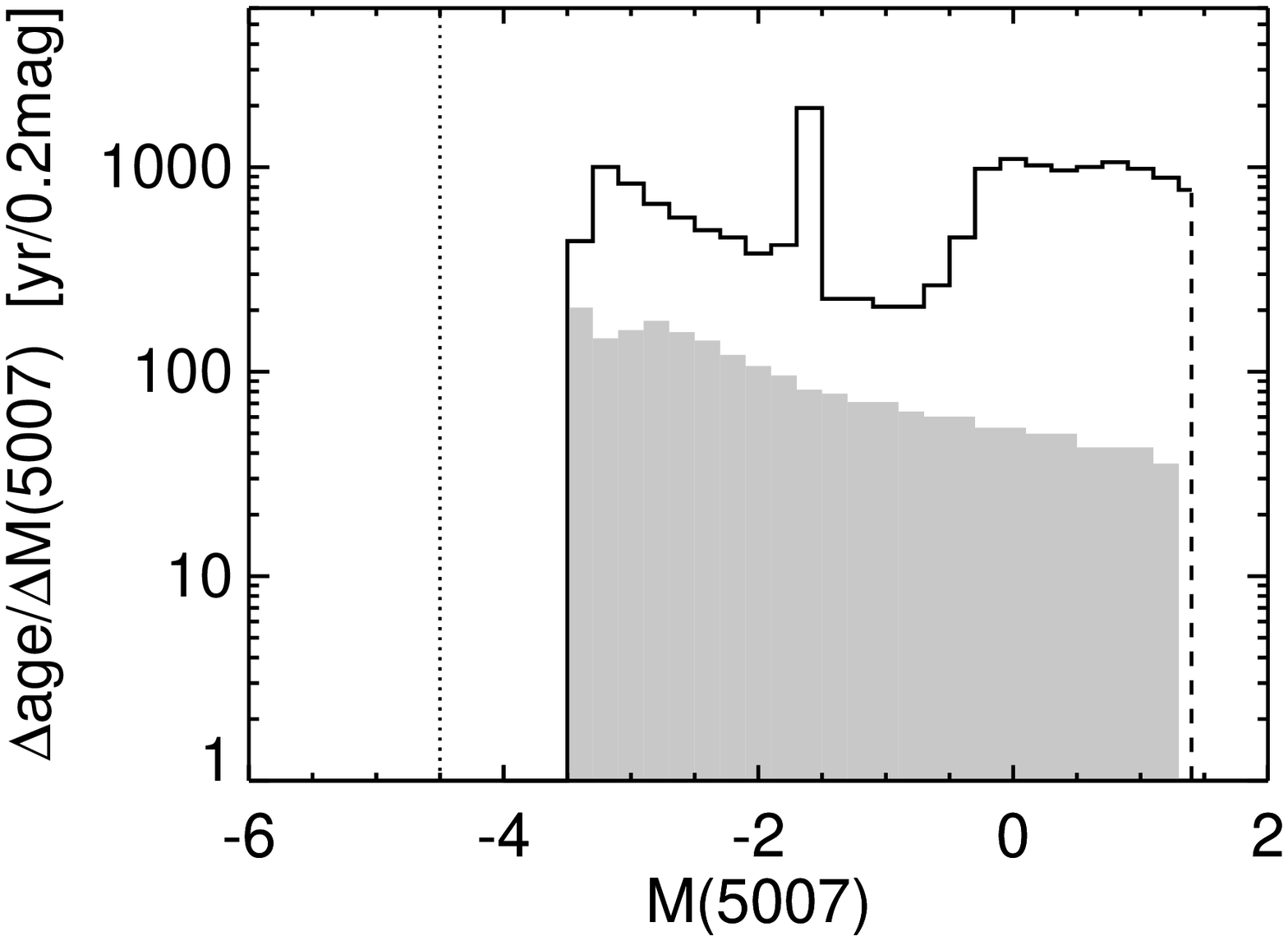}
}\\ \mbox{
\includegraphics[width=0.5\columnwidth]{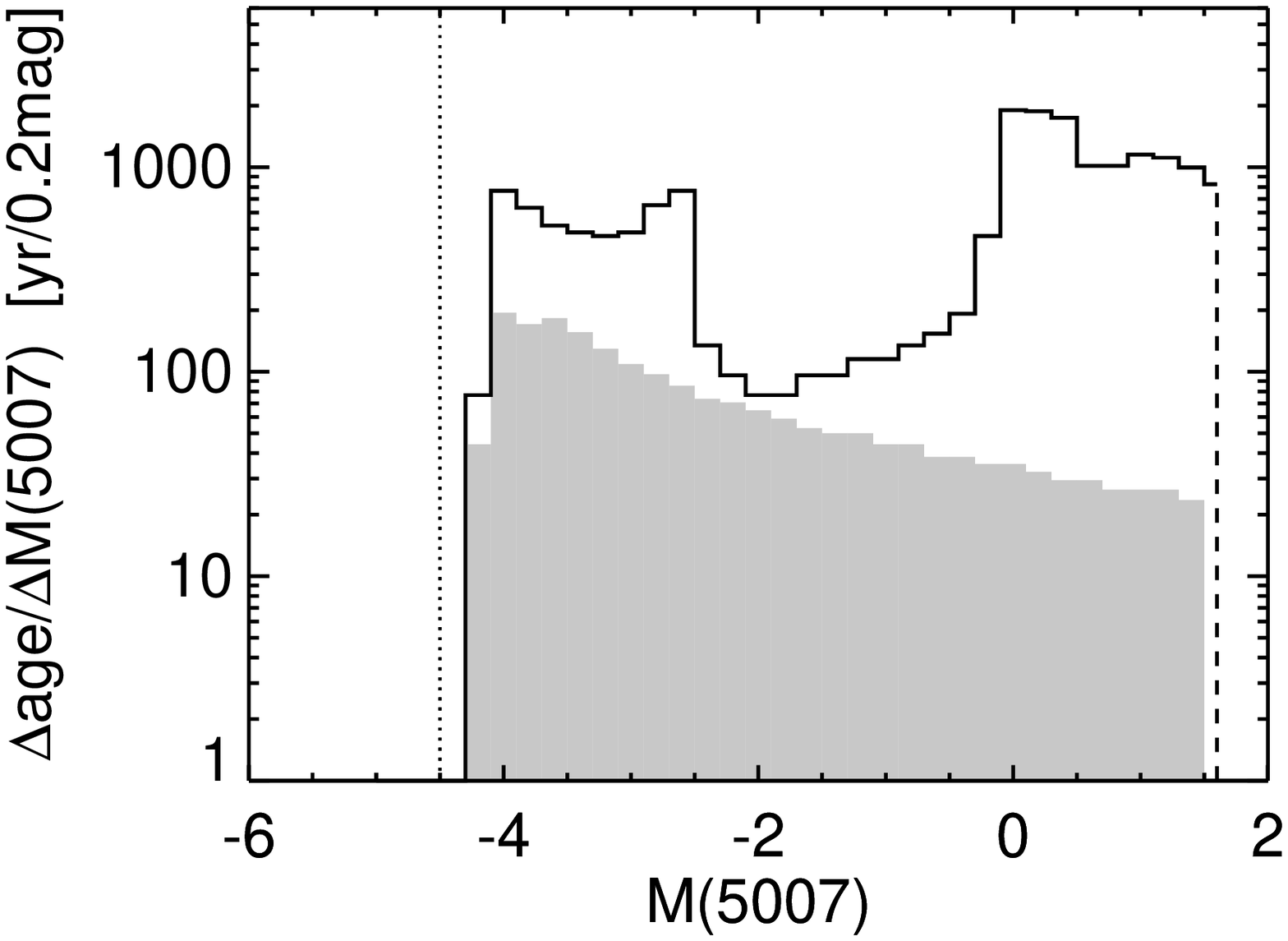}
\includegraphics[width=0.5\columnwidth]{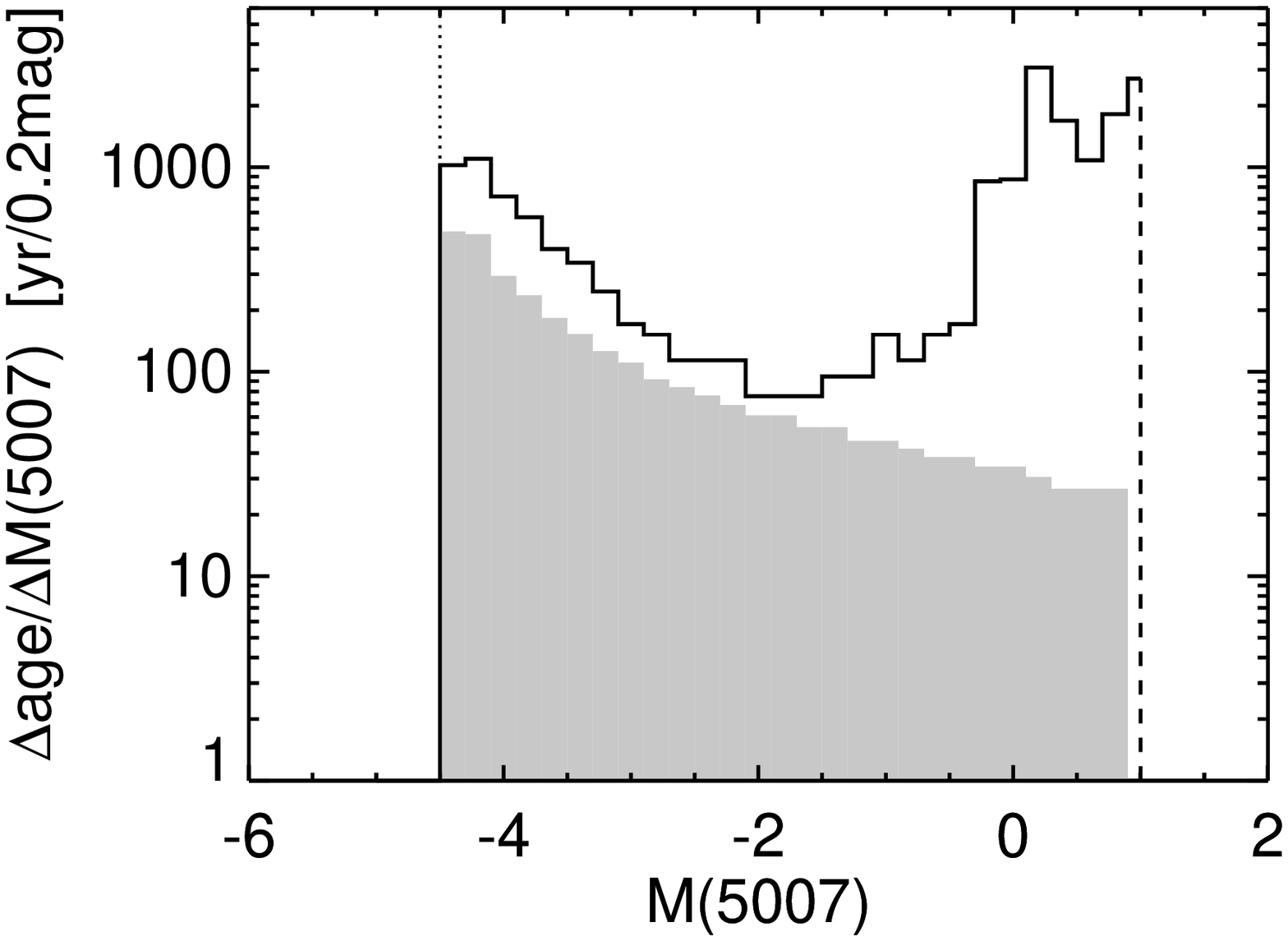}
}\\  \mbox{
\includegraphics[width=0.5\columnwidth]{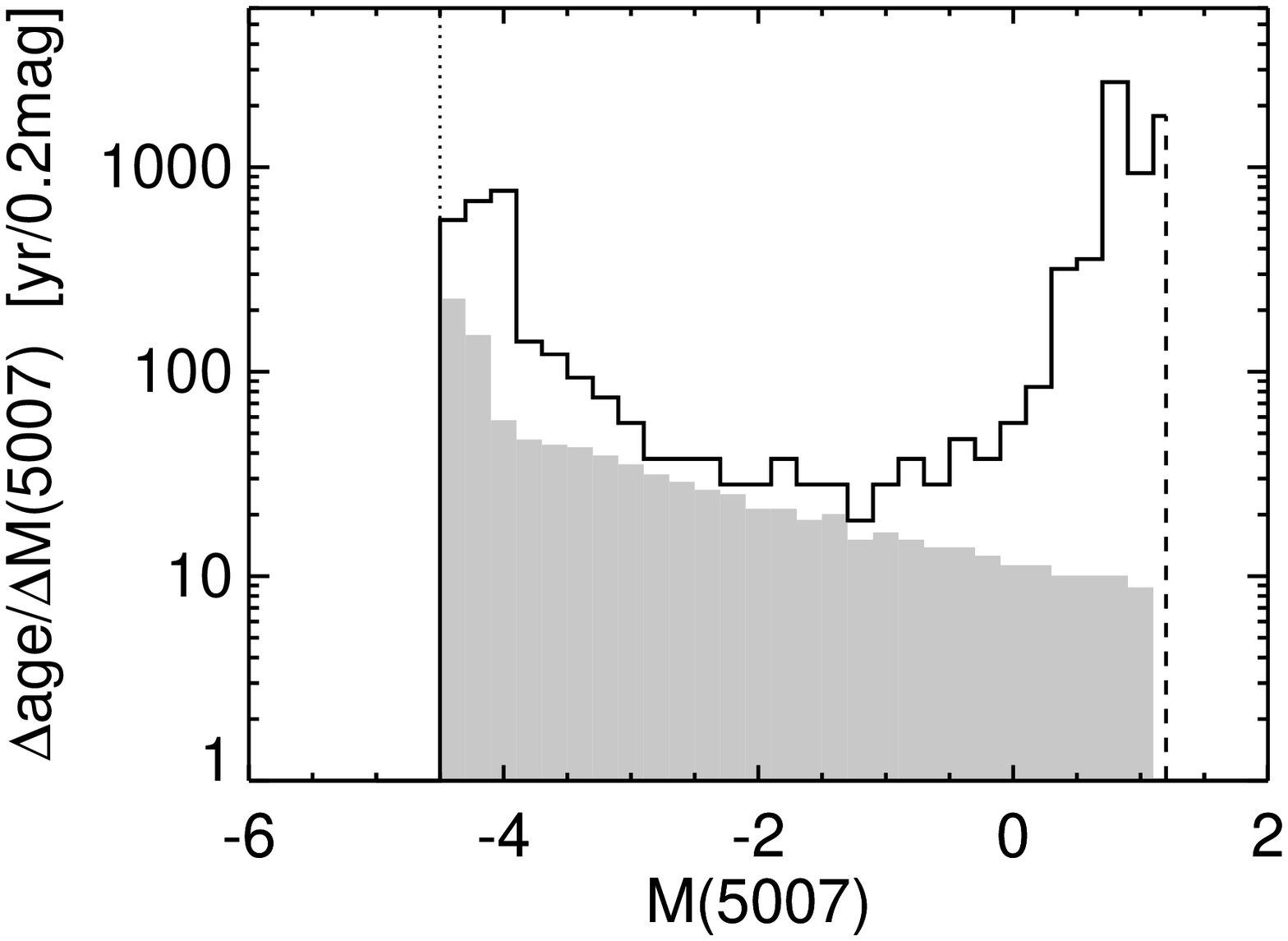}
\includegraphics[width=0.5\columnwidth]{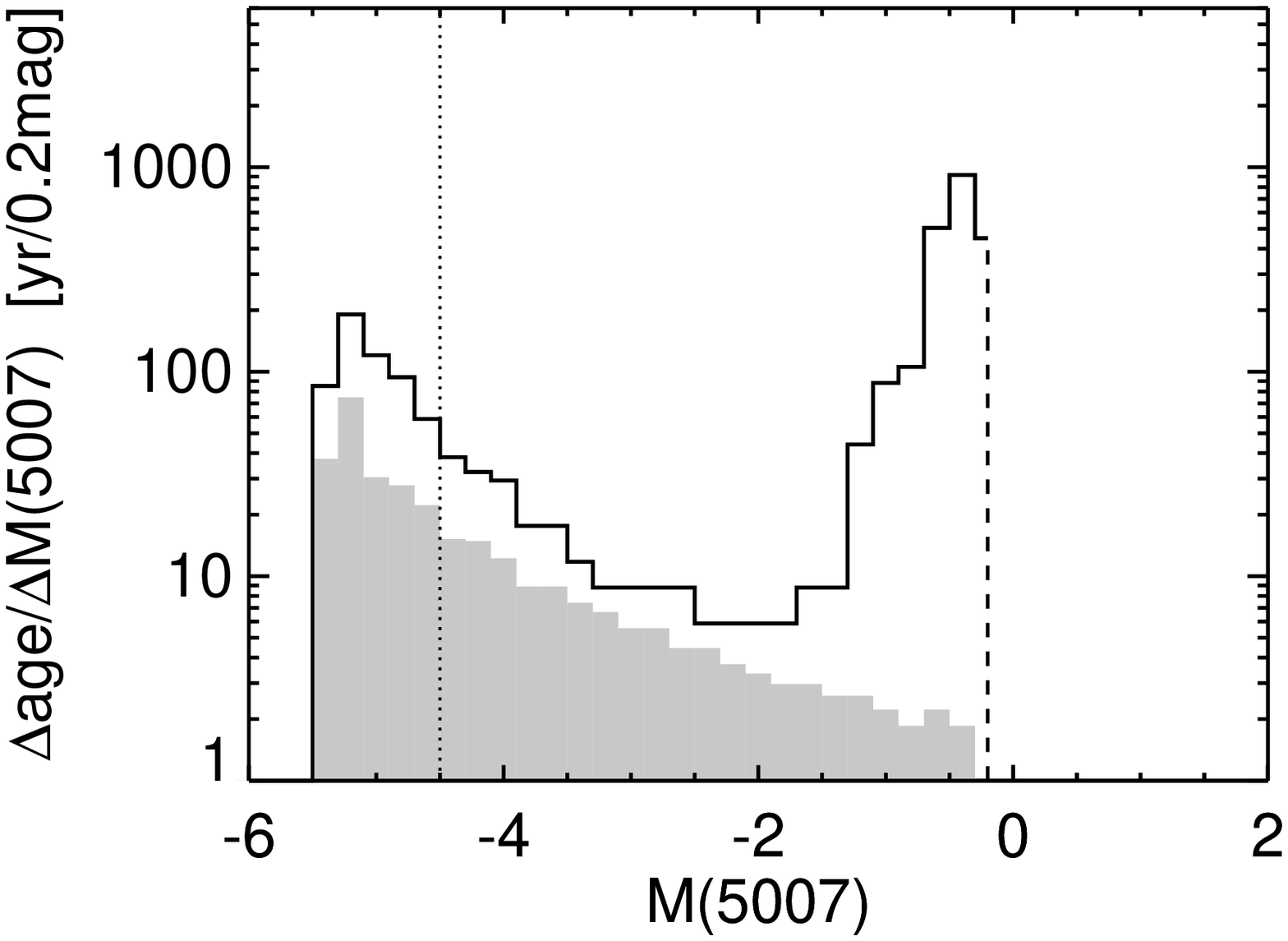}
}
\caption
{Histograms of the luminosity function proxy, $\Delta age/\Delta M(5007)$, for
 different simulations with central stars of 0.565\,\Msun\ (sequence No.\,22,
 \emph{top left}), 0.595\,\Msun\ (sequence No.\,6a, \emph{top right}), 0.605\,\Msun\
 (sequence No.\,6, \emph{middle left}), 0.605\,\Msun\ (sequence No.\,4,
 \emph{middle right}), 0.625\,\Msun\ (sequence No.\,8, \emph{bottom left})
 and 0.696\,\Msun\ (sequence No.\,10, \emph{bottom right}).
 Plotted are the total times spent
 per  absolute magnitude bin of 0.20~mag.   The gray histograms illustrate the
 contributions from increasing line brightness.  The dotted
 vertical line marks the observed bright $M(5007)$ limit.  The faint end of the
 different luminosity functions, indicated by vertical dashed lines, are affected
 by the limited simulation times (cf. Fig.\,\ref{M5007}).
}
\label{log.PNLF}
\end{figure}

  In passing we note that our simulations predict also a lower limit of the PNLF
  at about \mbox{$M(5007)\simeq 2$}\,mag.   The models at this limit are characterised by
  white-dwarf central stars with $\simeq$\,200~\Lsun\ and slowly re-ionising nebulae
  (Fig.\,\ref{M5007}, bottom panel).

  The observed luminosity functions are usually plotted on a logarithmic scale.
  We thus present in Fig.~\ref{log.PNLF} our individual luminosity functions from
  Figs.\,\ref{605.LF} and \ref{rest.LF} in a logarithmic form.
  Assuming that the mass distribution of central stars is very narrow and peaked at about
  0.6~\Msun, our simulations predict a dip in the PNLF around $M(5007)\simeq -2$\, mag.
  This dip is, of course, the signature of the extremely rapid fading of hydrogen-burning
  central stars once they have passed their maximum effective temperature.

\section{The \hb\ luminosity functions}
\label{hbeta.pnlf}

  For completeness we discuss here the luminosity functions based on the \hb\ line.
  This recombination line has the advantage that it depends only weakly on the
  electron temperature and is not afflicted by possible abundance variations like
  the collisionally excited lines of heavy ions, e.g.\ oxygen.
  The disadvantage is that \hb\ is usually weaker than \oiii\ which limits
  its use in extragalactic research.

\begin{figure}[t]             
\includegraphics[bb= 2.1cm 2.1cm 19.5cm 15.3cm, angle= -90, width=0.98\columnwidth]{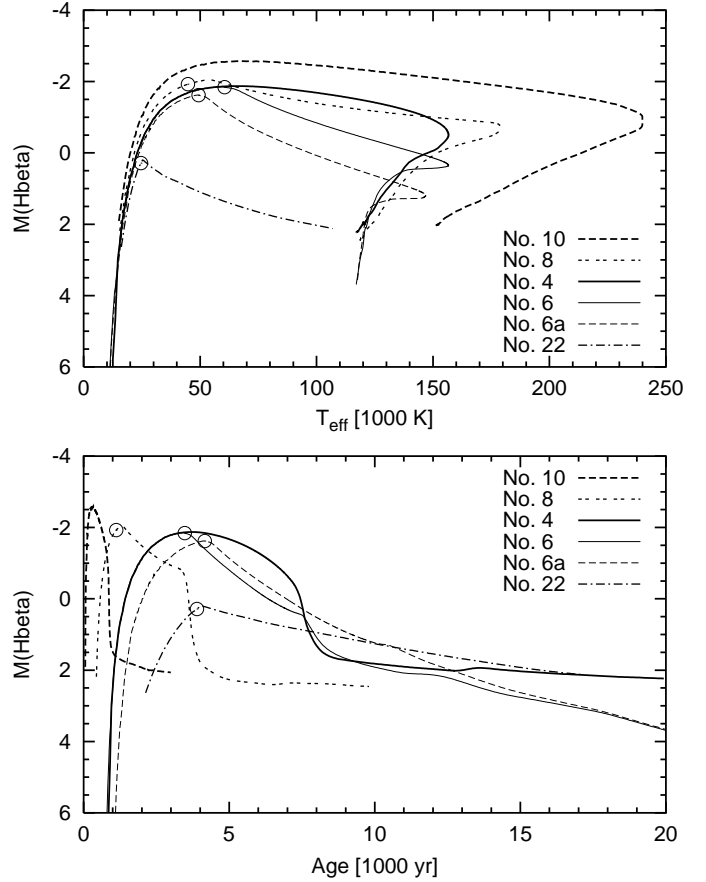}
\caption
{Brightness evolution of the \hb\ line for the sequences listed in Table~\ref{tab.mod}
         vs.\ central-star effective temperature (\emph{top}) and post-AGB age
	 (\emph{bottom}).  The circles indicate the
	 moments when the model nebulae become optically thin in the hydrogen
	 Lyman continuum.  Note that the sequences Nos.\ 4 and 10 remain
	 optically thick during their whole computed evolution, while the
	 nebular models of sequence No.\ 22 turn into the optically thin stage
	 already at a rather low effective temperature.
}
\label{Mumag.beta}
\end{figure}

  Figure\,\ref{Mumag.beta} is the same as Fig.\,\ref{M5007}, but for \hb.  As in the
  \oiii\ case the brightness increase is rapid while the following decrease occurs
  more gradually.  Of course, the peaks due to recombination as seen in
  Fig.\,\ref{M5007} cannot occur for hydrogen.

\begin{figure}[t]            
\mbox{
\includegraphics[width=0.5\columnwidth]{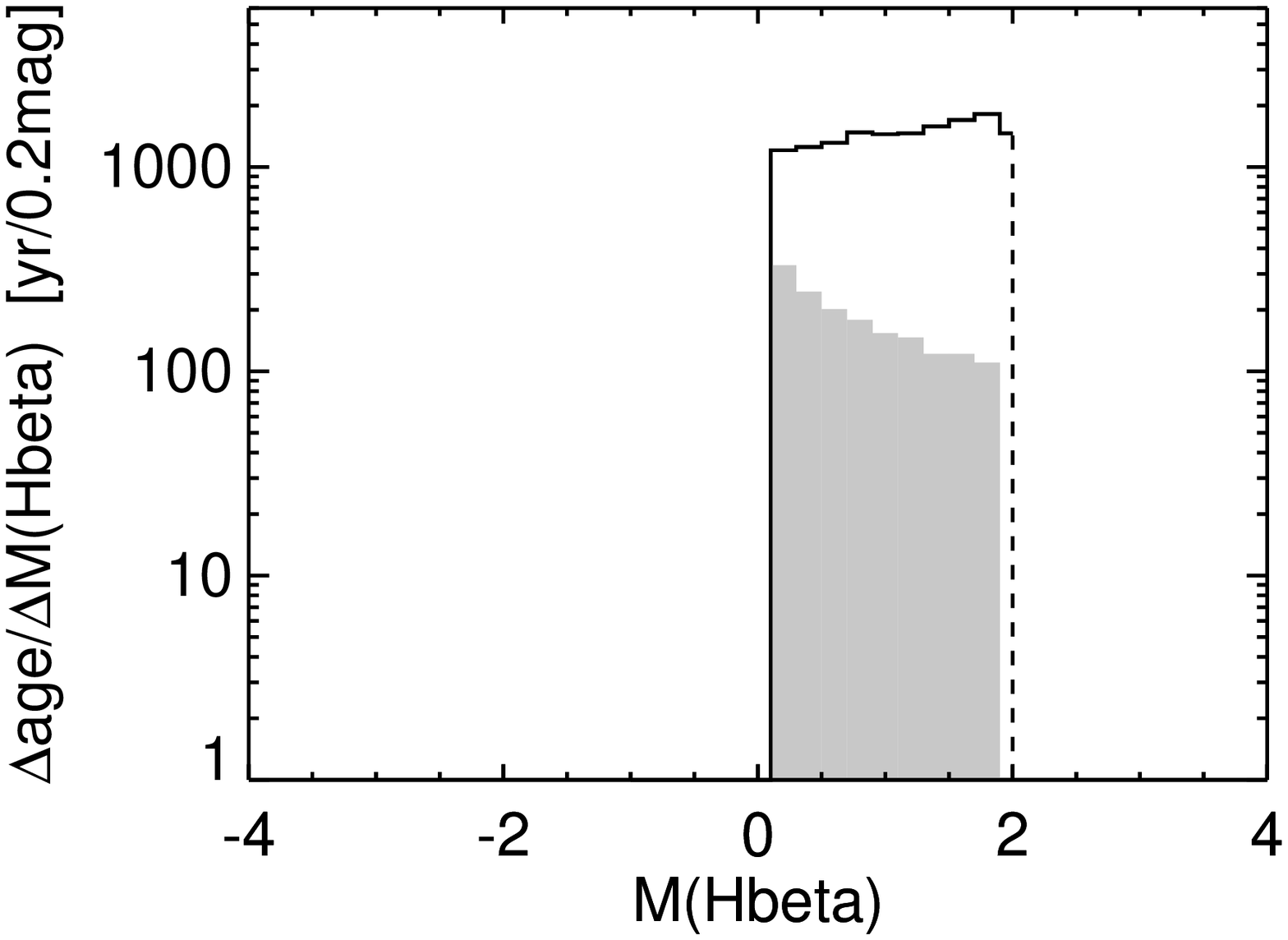}
\includegraphics[width=0.5\columnwidth]{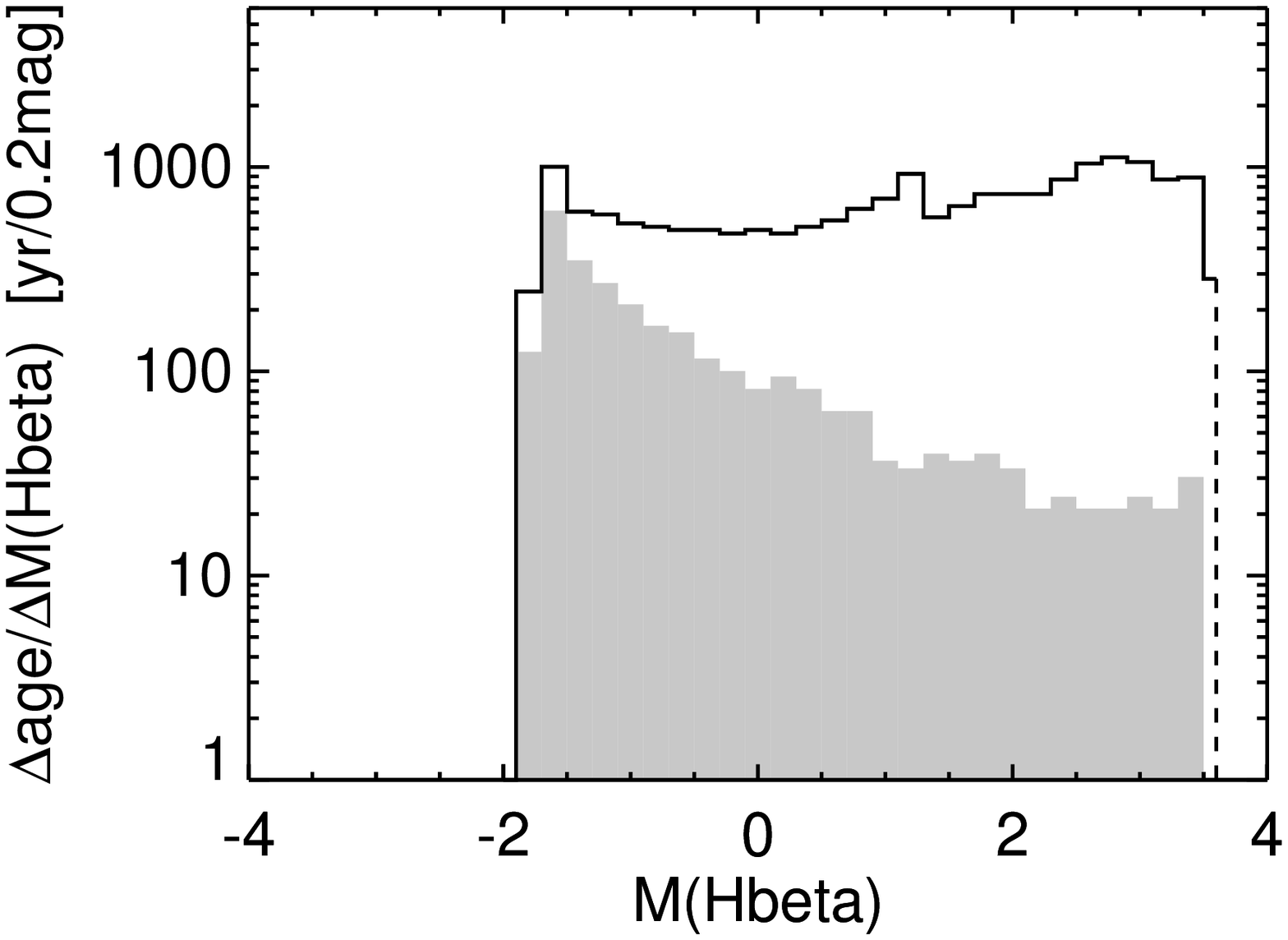}
}\\ \mbox{
\includegraphics[width=0.5\columnwidth]{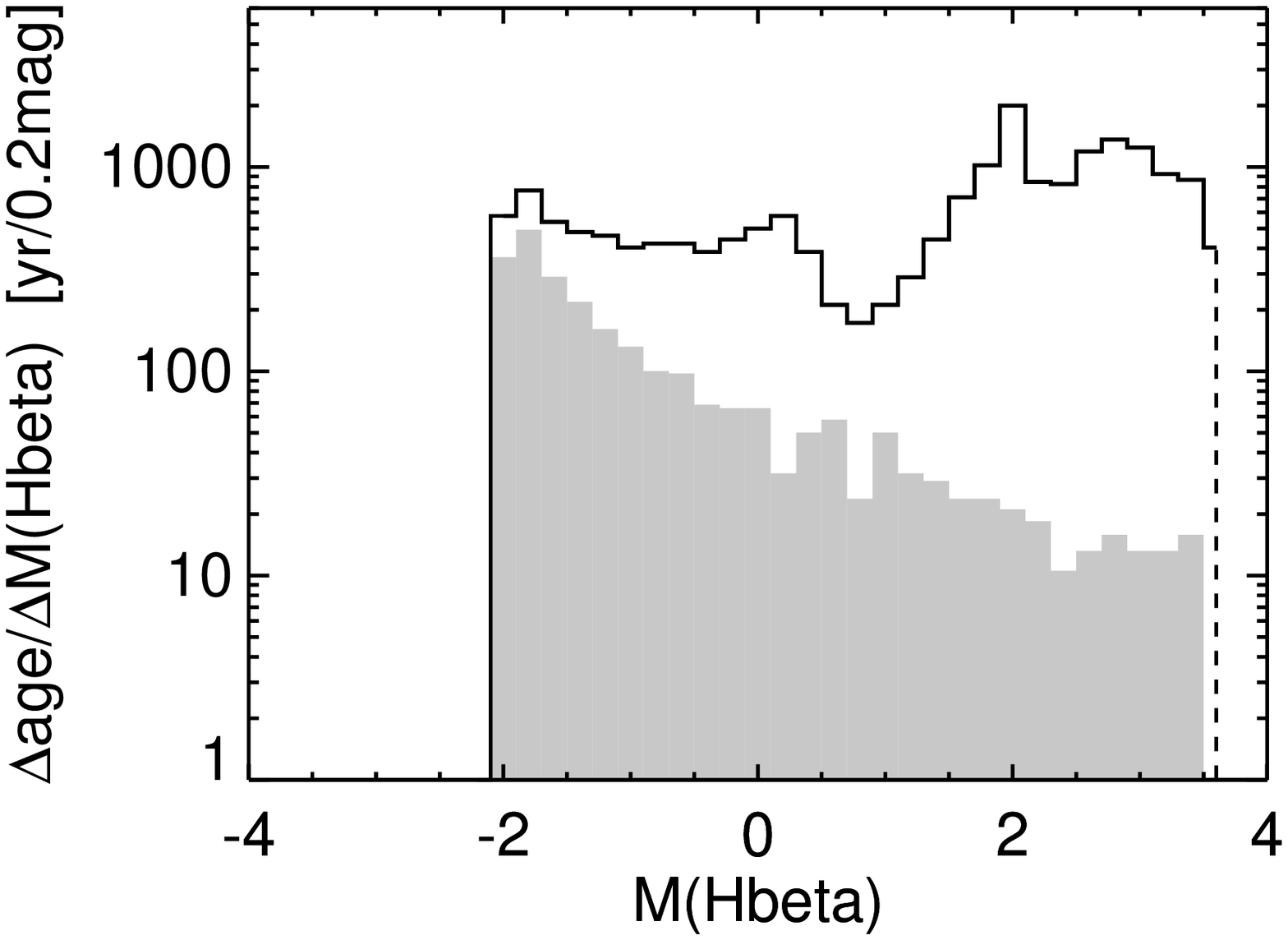}
\includegraphics[width=0.5\columnwidth]{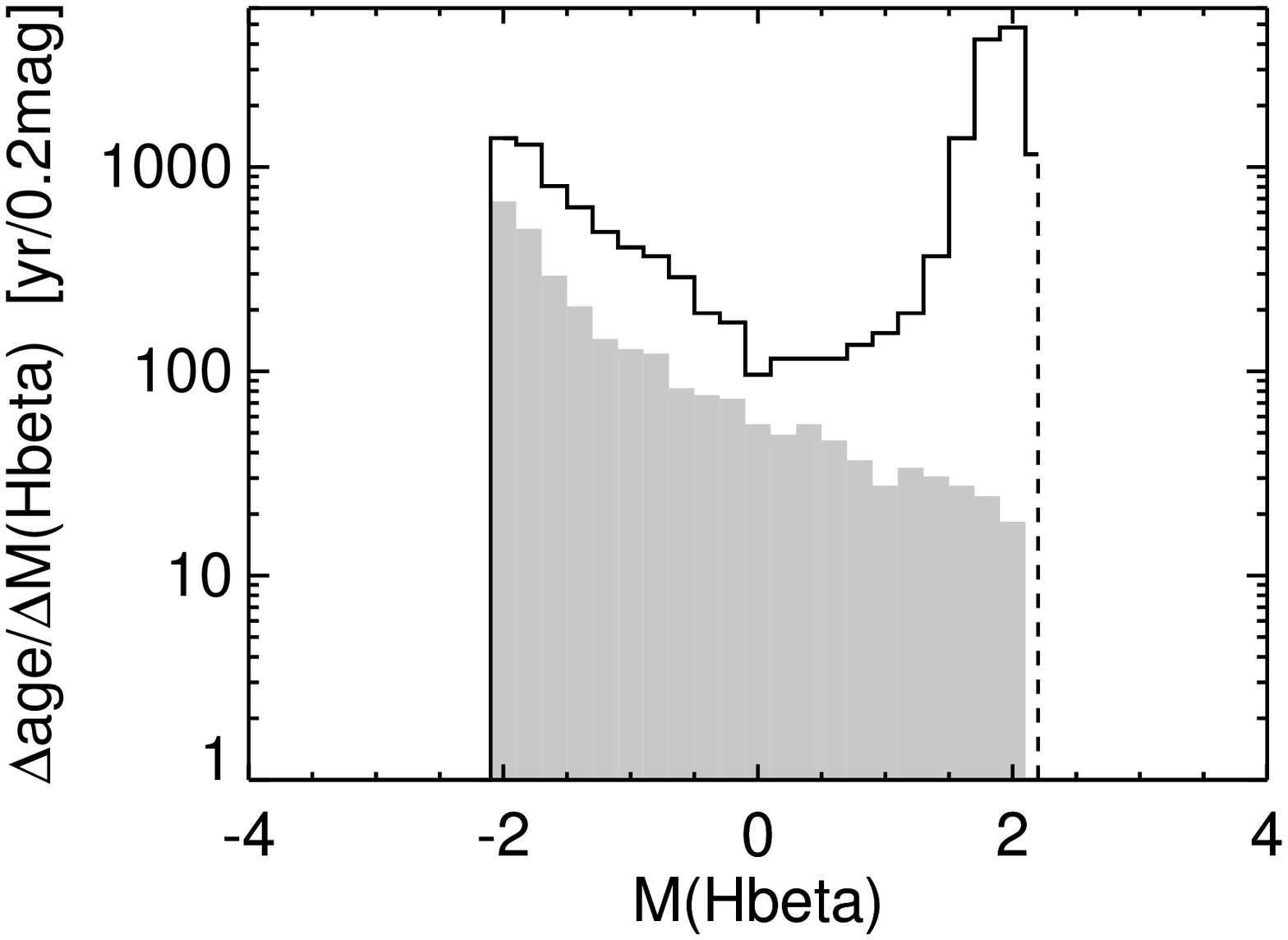}
}\\ \mbox{
\includegraphics[width=0.5\columnwidth]{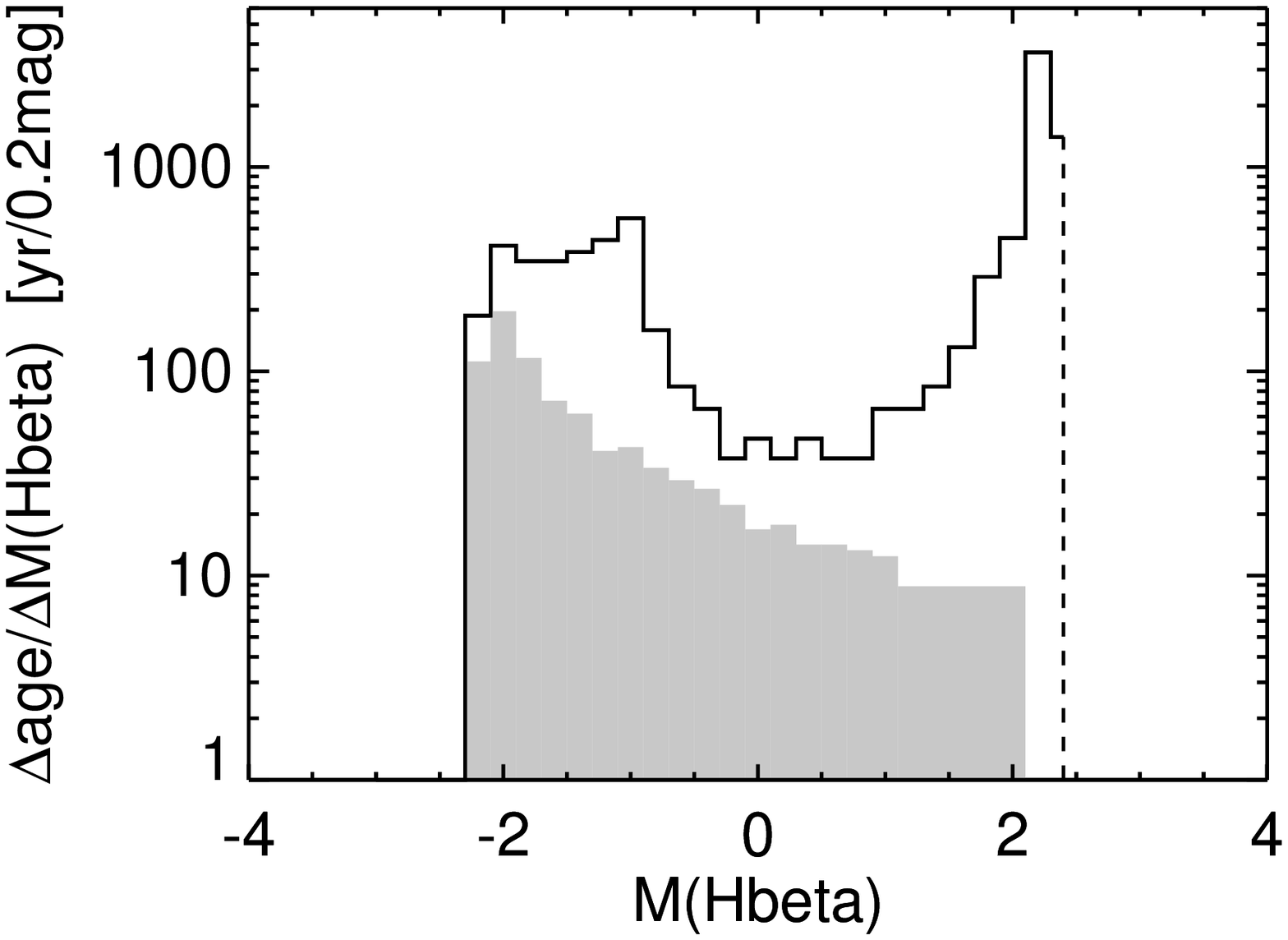}
\includegraphics[width=0.5\columnwidth]{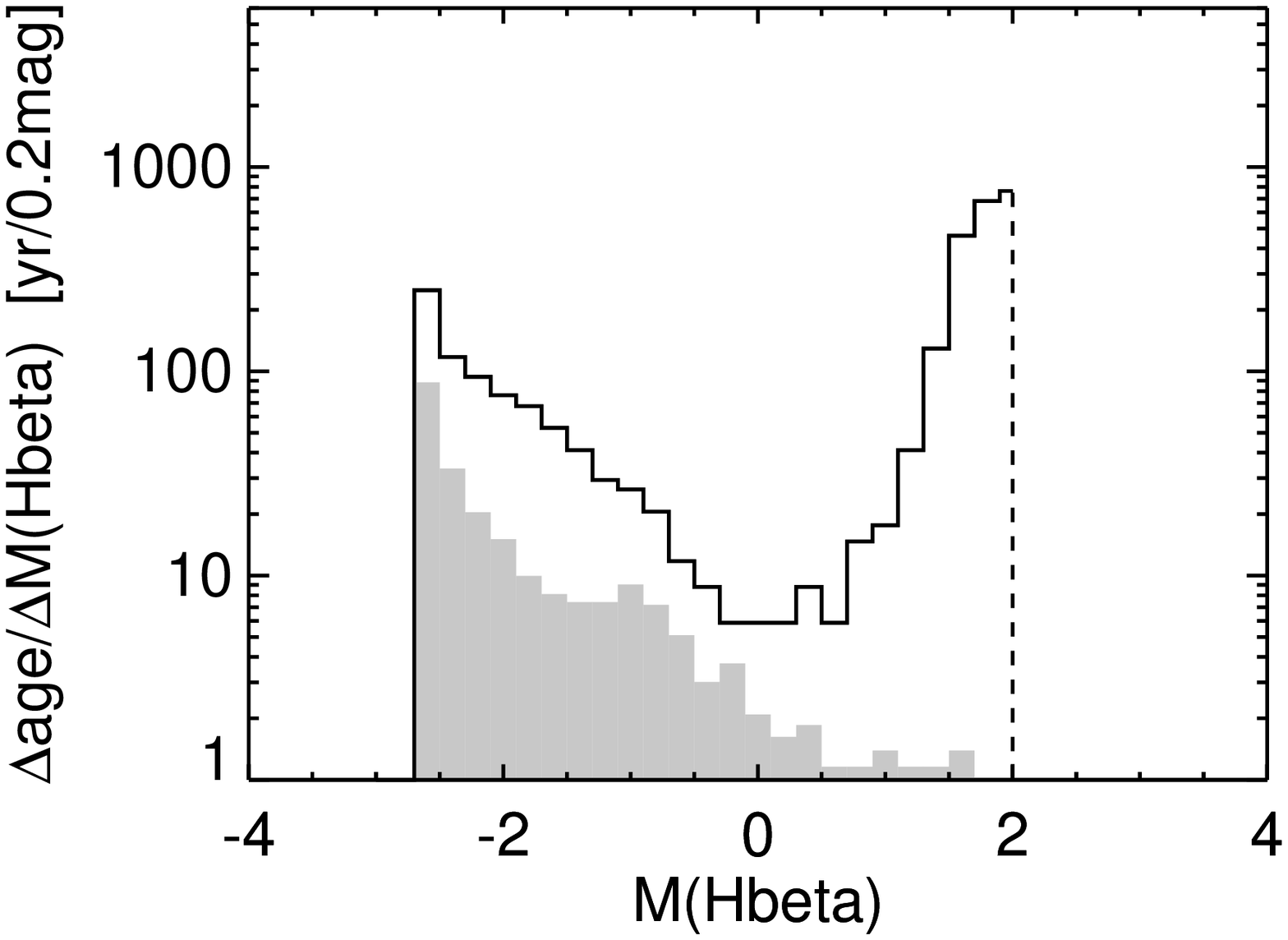}
}
\caption
{Same as in Fig.\,\ref{log.PNLF} but for $M({\rm H}\beta)$.
}
\label{Hbeta.PNLF}
\end{figure}

  The brightness evolution of \hb\ is converted into the luminosity function proxy,
  $\Delta age/\Delta M({\rm H}\beta)$, and the results are shown in
  Fig.\,\ref{Hbeta.PNLF}.  By comparison with Fig.~\ref{log.PNLF} we estimate that
  $M^{\star}(5007)= -4.5$
  corresponds to an \hb\ cut-off brightness of $-2.2\ldots-2.1$\,mag.
  Although the \hb\ luminosity function is fainter by about 2.5 magnitudes, it has
  also advantages: it does not directly depend on metallicity.  The dependence on
  the electron temperature is weak, thus also the indirect influence of the
  metallicity is expected to be modest.  The dip due to the fast dimming of the
  central star is expected at about 0.5 mag.

\section{Discussion}                        
 \label{dis}
\paragraph{Summary}
  With a set of 1D radiation-hydrodynamics simulations aimed at a better
  understanding of the formation and evolution of planetary nebulae we were able to
  investigate how the emission of important lines varies during the central-star's
  evolution towards a white-dwarf configuration and how they depend on the different
  combinations of central-star mass and initial envelope structure.
  Since the most effective conversion of stellar radiation
  into line emission occurs only for optically thick shells,
  it is of paramount importance for the interpretation
  of luminosity functions to know when and at which evolutionary phase PNe
  become optically thin or whether they will remain  always optically
  thick.

  Based on our simulations with hydrogen-burning post-AGB stellar models we found
  in particular:
\begin{itemize}
\item
  The optical depth in the Lyman continuum depends sensitively on the detailed
  nebular density structure and, of course, on the expansion speed.  The change of
  the optical depth with time is also controlled by the time evolution of the
  stellar parameters.
  Nebulae around central stars with masses $\la 0.6$~\Msun\ become optically thin
  in the Lyman continuum during the early part of the horizontal evolution across
  the Hertzsprung-Russell diagram \hbox{($\teff< 50\,000~{\rm K}$)}, in good agreement
  with the observational facts \citep[see][]{MKH.92}.  Central stars slightly
  more massive, however, evolve fast enough so that there is a large probability
  that their nebulae remain optically thick, or at least partially thick, during
  the high-luminosity part of evolution.  For masses $\ga\!0.65$~\Msun\ the star
  evolves already so quickly that it is safe to assume that the nebular shell will
  never become transparent for UV photons, except maybe when the nebular matter
  becomes very extended and merges with the surrounding interstellar medium.
  An example is \object{NGC\,7027} which is at present optically thick and has
  probably never been optically thin before
  \citepalias[see discussion in][]{Schetal.05a}.
\item
  With the oxygen abundance typical for galactic disk PNe, the observed \oiii\ cut-off
  luminosity corresponds to a stellar mass of 0.62~\Msun, provided the PN remains
  optically thick.  If the envelope is thin, the stellar luminosities, and hence the
  masses, must be larger accordingly.
  While most PNe are certainly optically thin, the ones that populate the bright
  end of the PNLF are optically thick, at least partially, with central stars
  slightly more massive than 0.6~\Msun.  Support for this interpretation comes
  from a PNe sample of the Magellanic Clouds for which we demonstrated that
  the excitations of the brightest objects in \hb\ and \oiii\ are exactly explained
  by optically thick (or only marginally thin) nebular shells around central stars with
  masses slightly in excess of 0.6~\Msun.
\end{itemize}

  The properties of our  hydrodynamical models are at variance with those
  of \citet{Ma.01, Ma.04}.  Their models attain a
  maximum \oiii\ luminosity when they become optically thick during the stellar
  luminosity decline well beyond effective temperatures of 100\,000~K.
  Since the conversion of UV energy into \oiii\ line emission is there less
  effective, and since also the stars are well beyond their maximum luminosity,
  \cite{Ma.04}   state that quite large central-star masses,
  $\approx\! 0.70\dots 0.75$~\Msun, are necessary in order to reach
   the observed cut-off line luminosity.

  We demonstrated, however, that our models are superior to the models used by
  \citet{Ma.01, Ma.04} because the latter
  {suffer from shortcomings based on a too a simple physical description}.
  The success of our models is based on the fact that the
  hydrodynamical simulations take proper care of the interplay between heating
  and expansion due to photoionisation by the radiation field and compression by
  wind interaction.   The dynamical effect of
  photoionisation is of paramount importance because it drives a shock wave through
  the surrounding AGB material, creating thereby a shell with a high-density outer
  edge which delays the break-through of the \ion{H}{i/ii} ionisation front for some time.
  Only during the further evolution the
  inner part of the shell is being compressed into the rim due to the increasing
  pressure exerted by the hot bubble which is shock-heated by the central-star wind.
  The shell contains most of the nebular mass and determines
  the size of the PN and its expansion rate.
  Any attempts to replace the adequate hydrodynamical treatment of these processes
  by simpler methods must fail.

\paragraph{Helium burning Wolf-Rayet central stars}
  The present study rests entirely on nebular models around hydrogen-burning
  post-AGB models.  It is expected that any sample of planetary nebulae is
  contaminated by objects whose central stars are burning helium while they
  evolve off the AGB, e.g. by objects with (carbon-rich) Wolf-Rayet central stars
  with strong stellar emission lines and hydrogen-depleted surfaces.
  Observationally, they may make up for about 25--30\,\% of the total PNe population.
  However, \citet{GSEC.04} report a smaller fraction of only about 10\,\%.
  Because a convincing theory on the formation and evolution of PNe with
  Wolf-Rayet central stars is still missing, one can only estimate their possible
  impact on the PNLF.

  Assuming that their mass distribution is about the same as that of the hydrogen-rich
  central stars, we may use the results of \citet{VW.94} for helium-burning
  central stars to estimate the expected behaviour of
  the line emission\footnote{Note that these models still have hydrogen-rich envelopes.}.
  The evolution of helium-burning models differs twofold from that
  of the hydrogen-burning models discussed here:
\begin{enumerate}
\item  They evolve more slowly through the PN domain which will favour optically
       thin nebulae.
\item  They do not cross the HR diagram with a virtually constant luminosity.
       Instead, their luminosity decreases such that at
       \mbox{$\teff \simeq 100\,000$~K} they are less luminous
       than hydrogen-burning models of the same mass by about a
       factor of 2 to 3\,\footnote{The upturn shown by the \citeauthor{VW.94} models is
       due to the rekindling of hydrogen and will \emph{not} occur in objects with
       hydrogen-depleted envelopes.}.
       Because of the lower stellar luminosity, a PN with a Wolf-Rayet nucleus of
       about 0.6~\Msun\ \emph{is not expected to contribute to the \oiii\ cut-off
       luminosity, even if its nebula shell would be optically thick.}
\end{enumerate}

  There exists the possibility that hydrogen-burning central stars turn into
  helium-burning ones by a thermal pulse
  occurring during the transit to the white-dwarf region.  Such a pulse forces the
  star to return briefly back to the vicinity of the AGB.  During this transient
  evolution the stellar surface may become hydrogen-free
  by mixing and burning.  More details about this so-called `born-again' scenario
  can be found in \cite{He.01}.   The likelihood of such late thermal pulses is, however,
  very low and need not be considered here.

  We note in passing that the sample of Galactic PNe used in Fig.\,\ref{excite.mag}
  contains only nuclei with hydrogen-rich surfaces which are believed to burn hydrogen
  in a shell.  In contrast, the fraction of hydrogen-depleted [WC] central stars for
  the LMC sample presented in Fig.\,\ref{excite.LMC} is not known.

\paragraph{The core-mass luminosity relation}
  We remind the reader that it is the luminosity of a PN nucleus which can only be
  determined and which is responsible for the PN's line emission.  Reliable
  masses are only available for a few cases, but in general they are derived by
  converting luminosity into a (post-AGB) mass using a core-mass luminosity relation
  based on canonical evolution theory.  This relation was originally found by
  \citet{P.70}, and our post-AGB models provide a relationship very close to that of
  \citeauthor{P.70}.

  The relationship between luminosity and core mass, or virtually total mass for
  post-AGB objects, depends also somewhat on the metallicity.  
  The only available set of post-AGB models to date computed for different metallicities
  is that of \citet{VW.94}.  A comparison with our models shows that the core-mass
  luminosity relation of their solar-metallicity models agrees reasonably well with ours.
  The most metal-poor models \hbox{($Z=0.001$)} are, however, fainter by
  $\Delta\log L \simeq 0.05$ in the important region between 0.6 and 0.65~\Msun.

  Important is that AGB models which include convective overshoot as introduced by
  \citet{He.97} show a different behaviour between size and mass of the core as compared
  to standard evolutionary calculations, resulting also in a different relation between
  core-mass and luminosity.
  In particular, an unique relation between core mass and luminosity cannot be derived, and
  as shown by \citet{He.98}, the stellar luminosity may exceed 
  the one predicted by the canonical evolutionary theory in which convective overshoot
  is ignored.   As a consequence, the luminosity of a post-AGB star of given mass may
  be larger than predicted by Paczy\'nski's law, but by how much depends in a
  complicated manner on the dredge-up and mass-loss history on the AGB and cannot
  presently be answered because of our rather limited understanding of stellar
  convection.   Judging from \citet{He.98} it appears plausible that the masses
  used in our model simulations could be too large by a few 0.01\,\Msun\ at most.

\paragraph{The universal \oiii\ cut-off luminosity}
   It is still an enigma why the bright cut-off magnitude of the PNLF is virtually
   independent of the properties of the parent stellar population like, e.g.,
   metallicity or turn-off mass.  With our limited set of model simulations we are
   certainly not in a position to solve this problem.  A detailed
   knowledge of mass loss and dredge-up along the AGB, as a function of the stellar
   parameter and the metallicity, is needed for any progress.
  
   One should also consider that
   the maximum of the conversion efficiency of the \oiii\ 5007 \AA\ line decreases
   with metallicity and shifts also to lower effective temperatures
   \cite[cf. Fig. 2 in][]{DJV.92}.  For instance, at one tenth of the solar
   metallicity, the maximum 5007 \AA\ emission occurs at $\teff \simeq 80\,000$~K,
   with a conversion efficiency reduced by a factor of about two.  The reason is
   simple: at a lower oxygen abundance, O$^{+2}$ occurs as the main ionisation
   stage already at smaller photon fluxes, i.e. at lower stellar temperatures.
   The reduction of the oxygen abundance is partly compensated for by a larger
   electron temperature.

   It might well be that the combined influence of 
   all the factors discussed here  leads to a universal \oiii\ cut-off line luminosity.
   A satisfying answer, however, is only possible with many more simulations of the
   kind presented here, with all the
   discussed issues considered, and is far beyond the present work.

\begin{acknowledgement}
This work benefitted from discussions with A. Teodoresco and R. M\'endez.
C.S acknowledges support by DFG grant SCHO 394/26.
\end{acknowledgement}

\end{document}